\documentclass[aps, prb, twocolumn]{revtex4-2}
\usepackage{bm}
\usepackage{graphicx}
\usepackage{dcolumn}
\usepackage{bm}
\usepackage{color,hyperref}
\hypersetup{colorlinks=true, linkcolor=blue, citecolor=blue, urlcolor=blue} 
\usepackage{physics}
\usepackage{amsmath,amssymb}
\usepackage{braket}
\usepackage{xcolor}
\usepackage{soul}

\usepackage{tikz}
\usetikzlibrary{arrows.meta,positioning,decorations.pathreplacing}

\begin{document}
\title{Operator space transport in collective spin Lindbladians}
\author{Dominik Nemeth}
\email{dominik.nemeth@manchester.ac.uk}
\author{Ahsan Nazir}
\author{Robert-Jan Slager}
\author{Alessandro Principi}

\affiliation{Department of Physics and Astronomy, The University of Manchester, Oxford Road, Manchester M13 9PL, United Kingdom}

\date{\today}

\begin{abstract}
Boundary time crystals (BTCs) are paradigmatic examples of continuous time crystals that emerge in collective, dissipative quantum systems. Their microscopic origin, however, remains obscured by the complexity of Lindbladian many-body dynamics and has largely been investigated through semiclassical and numerical approaches. Here, we develop an exact operator space framework for collective spin Lindbladians based on irreducible tensor representations. We map the Lindbladian dynamics onto a local non-Hermitian hopping problem, where weak symmetries impose selection rules on operator transport, while the hopping amplitudes and their relative strengths determine the resulting dynamical regimes. We identify the microscopic ingredients that distinguish models supporting conventional collective precession from those supporting BTC dynamics. Our approach further reveals how non-unital dissipation can provide a source for the traceless operator subspace, providing a microscopic mechanism for the reduced sensitivity of BTC dynamics to initial conditions beyond mean-field descriptions. More broadly, our results establish operator space transport as a general perspective for understanding dissipative many-body dynamics and its connections to non-Hermitian phenomena.
\end{abstract}

\maketitle

\section{Introduction}
Open quantum many-body dynamics has become a central arena for uncovering novel phases of matter beyond equilibrium, a particular example of which is a time crystal phase. While time crystals were originally proposed in closed, periodically driven systems \cite{Wilczek2012,Bruno2013, Watanabe2015, Khemani2016, Khemani2019, Sacha2015, Sacha2018, Else2016, Else2017, Else2020, Keyserlingk2016, Yao2017, Lazarides2017, Zaletel2023, Zhang2017}, their extension to open quantum systems has revealed new mechanisms by which persistent oscillations can arise under dissipation in macroscopic systems \cite{Seibold2020, Lledo2020, Booker2020, Passarelli2022}. A paradigmatic example is the boundary time crystal (BTC)~\cite{Iemini2018}, which emerges in macroscopically large collective spin systems forming the boundary of a much larger, weakly coupled Markovian environment.

Early studies of boundary time crystals have largely relied on semiclassical and mean-field descriptions~\cite{Iemini2018, Piccito2021}, where the dynamics is reduced to nonlinear equations of motion and time crystal behavior is identified through linear stability analysis. However, the validity of such mean-field approximations in open many-body systems has been actively debated~\cite{Mori2013, Carollo2021, Fiorelli2023, Mukherjee2024}. Recent works that go beyond mean-field theory~\cite{Carollo2022, Mukherjee2024, Nakanishi2023, Nemethb2025, Liu2025} have revealed that BTCs possess genuinely many-body features that are not fully captured within semiclassical descriptions.

These developments highlight the need for a microscopic, fully quantum framework capable of describing collective spin Lindbladians beyond mean-field approximations. In particular, such a framework should allow one to systematically classify the resulting dynamical regimes and provide a natural basis for analytical calculations, thereby offering deeper insight into the origin and properties of BTC behavior.

In problems involving spin systems or angular momentum degrees of freedom, it is customary to work in the angular momentum eigenbasis, expanding operators in terms of matrix elements between states $\ket{j\, m}$, where $j$ and $m$ denote the total angular momentum and its projection along the $z$-axis, respectively. However, an alternative and often more physically transparent approach---ubiquitous in nuclear magnetic resonance (NMR)~\cite{Callaghan1993, Abragam1961, Slichter1996} and atomic, molecular, and optical (AMO) physics~\cite{Fano1957, Happer1972, Blum2012}---makes use of the spherical tensor operator formalism. In this representation, operators are expanded in terms of irreducible tensor operators $T^k_q$, characterized by tensor rank $k$ and component $q$, which transform under rotations via the adjoint action of the SU(2) group. For fixed $k$, these operators form a spin-$k$ multiplet~\cite{Varshalovich1988, Hecht2000}.

The dynamics of collective spin systems can be organized through couplings between angular momentum sectors~\cite{Wood2014, Ribeiro2019}. At the same time, several classes of open quantum systems admit a description directly in operator space. For example, in quadratic fermionic systems, the Liouvillian can be viewed as acting on a Fock space of operators and described through ladder superoperators~\cite{Prosen2008, Prosen2010}. This suggests the possibility of a similar framework for collective spin systems, motivating the question of whether Lindbladian collective spin dynamics can likewise be understood through structured couplings between irreducible operator subspaces.

In this work, we develop a microscopic quantum framework for classifying collective spin Lindbladians according to their weak symmetries, defined by superoperators that commute with the Lindbladian~\cite{Buca2012, Albert2014}. Unlike semiclassical approaches based on truncations of the hierarchy of correlations~\cite{Iemini2018, Piccito2021, Mukherjee2024}, our formulation retains the full operator space structure of the dynamics. We construct the framework using the irreducible decomposition of Liouville space, expanding the density operator in spherical tensor operators labelled by their rank $k$ and component $q$~\cite{Varshalovich1988, Hecht2000, Blum2012}. In this representation, the Lindblad master equation maps onto a local non-Hermitian hopping problem in operator space. Weak symmetries constrain the allowed directions of transport, while the corresponding microscopic hopping amplitudes determine the resulting dynamical regime. This provides a direct microscopic route to identifying the features that distinguish conventional collective spin precession from BTC dynamics.

The paper is organized as follows. In Sec.~\ref{sec:lme_collspin} we introduce the class of collective spin Lindbladians considered throughout. In Sec.~\ref{sec:representation_theory_for_collective_spins} we review the representation theory of collective spins and show how the relevant irreducible representations emerge in operator space. In Sec.~\ref{sec:classification_of_collective_spin_dynamics} we discuss collective spin models based on their weak symmetry constraints by expanding the density operator in the spherical tensor basis. We illustrate the utility of our framework by focusing on driven collective spin models that experience either collective dephasing or collective decay. Then, in Sec.~\ref{sec:initial_state_independent_dynamics} we discuss how non-unital dissipation can lead to a source term in operator space, which we use to explain the reduced initial state sensitivity of BTCs. We compare our method with semiclassical approaches in Sec.~\ref{sec:relation_to_semiclassical_approaches}. In Sec.~\ref{sec:perturbative_approach} we highlight how our framework can be used to study the extreme time crystal limit, which occurs in the strong-driving regime. In Sec.~\ref{sec:dissipation_driven_hybridization} we examine how increasing dissipation drives the hybridization of spherical tensor modes, which can be interpreted as a form of delocalization in operator space. Finally, we comment on the thermodynamic limit of the BTC model in Sec.~\ref{sec:thermodynamic_limit_btc_model}.

Throughout this work we employ natural units. The terms irreducible tensor operators and spherical tensor operators, as well as irreducible representations (irreps) and spherical tensor sectors, are used interchangeably.

\section{Collective spin dynamics under Markovian dissipation}
\label{sec:lme_collspin}
We consider open quantum systems governed by a Lindblad master equation for the density operator $\rho(t)$ ($\hbar=1$),
\begin{equation}
\begin{split}
    &\frac{d\rho}{dt} = \mathcal{L}[\rho]
= -i[H,\rho] + \mathcal D[\rho],\\
&\mathcal D[\rho] = \dfrac{\Gamma}{2j} \sum_\mu \Bigl(
L_\mu \rho L_\mu^\dagger
- \tfrac{1}{2}\{L_\mu^\dagger L_\mu,\rho\}
\Bigr), \\
\end{split}
\label{eqn:lindblad_me}
\end{equation}
where $\mathcal{L}$ is the Liouvillian superoperator generating Markovian evolution and $\mathcal D$ is the Lindblad dissipator. We consider collective spin models in which the Hamiltonian $H$ and jump operators $L_\mu$ are constructed from spin-$j$ operators $J_\alpha$ ($\alpha = x, y, z$), with a uniform collective rate $\Gamma/2j$. The factor $1/2j$ constitutes a Kac scaling and guarantees that the collective dissipative dynamics remain extensive as $j\to\infty$. These operators satisfy the $\mathfrak{su}(2)$ commutation relations $[J_\alpha, J_\beta] = i \epsilon_{\alpha \beta \gamma} J_\gamma$, with ladder operators $J_\pm = J_x \pm i J_y$. 

For the class of models considered here, the total spin $J^2$ is conserved, allowing each fixed-$j$ sector to be analyzed independently. Such models arise naturally in cavity systems with cooperative emission~\cite{Walls1978, Drummond1978, Drummond1980}, where $N$ spin-$\tfrac{1}{2}$ degrees of freedom behave collectively as a single spin with $j = N/2$.

In this work, we focus on the collective spin models defined by $H=\Omega J_x$ with either $L_\mu=J_z$ or $L_\mu=J_-$, which serve as minimal representatives of models, in which collective precession and BTC dynamics can emerge. In the precession case, the spin components $J_y$ and $J_z$ undergo single-frequency oscillations that generally depend on the initial state~\cite{Nakanishi2023, Nemethb2025}. This behavior arises from a simple Liouvillian structure dominated by a single oscillatory mode. In contrast, BTC dynamics exhibit qualitatively richer behavior: the oscillations display multiple frequency components and, crucially, exist independently of the initial state~\cite{Iemini2018, Piccito2021, Mukherjee2024}. In the thermodynamic limit ($j\to \infty$), the corresponding decay rate vanishes and these oscillations persist indefinitely, giving rise to the BTC phase~\cite{Iemini2018, Piccito2021, Nakanishi2023, Nemethb2025}.

The operators $J_\alpha$ generate a $\mathfrak{su}(2)$ Lie algebra, and their representation theory leads to the well known basis of angular momentum eigenstates $\{\ket{j \, m}\}$, with $m=-j,\ldots,j$. The dynamics can then be formulated in terms of a density matrix in the $\{\ket{j\,m}\}$ basis as,
\begin{equation}
    \rho = \sum_{m,m'} \rho_{mm'} \,\ket{j \, m} \bra{j \, m'},
    \label{eqn:rho_in_ang_mom_basis}
\end{equation}
for fixed $j$. This usual approach has been employed in Ref.~\cite{Nakanishi2023}, which performed a perturbative treatment of collective spin Lindbladians.

However, for open quantum systems governed by a Liouvillian superoperator, this state-based representation is no longer the most symmetry-adapted description. In the following section, we introduce an operator-based formulation using the adjoint representation of $\mathfrak{su}(2)$.

\section{Representation Theory for Collective Spins}
\label{sec:representation_theory_for_collective_spins}
In closed quantum systems the fundamental dynamical object is the state vector $\ket{\psi}\in\mathcal H$. For systems whose degrees of freedom furnish a representation of the 
$\mathfrak{su}(2)$ Lie algebra, which in the case of collective spins is generated by the operators $J_\alpha$, it is natural to adopt a state-based representation in which these generators act on $\mathcal H$. Representation theory then organizes the Hilbert space into irreps. Concretely, one diagonalizes the 
Casimir operator $J^2 = J_x^2 + J_y^2 + J_z^2$ (together with $J_z$), obtaining sectors labeled by total spin $j$ and magnetic quantum numbers $m = -j, \dots, j$, such that 
\begin{equation}
\mathcal H = \bigoplus_j \mathcal H_j,   \qquad \dim \mathcal H_j = 2j + 1.
\end{equation}
This decomposition provides the conventional state-based quantum numbers typically used to describe spin systems.

Conversely, in open quantum systems the fundamental dynamical object is the density operator $\rho$. In the collective spin case, $\rho$ belongs to the operator space $\mathcal B(\mathcal H_j)$, which is isomorphic to a tensor-product space. This can be seen from the fact that any operator may be constructed from the outer products of kets $\{\ket{j\,m}\}$ and bras $\{\bra{j\,m'}\}$. Following the Choi-Jamio\l{}kowski isomorphism (vectorization)~\cite{Choi1975,Jamiolkowski1972,Gyamfi2020}, one applies the mapping \mbox
{$
    \ket{j\,m}\bra{j\,m'} \;\mapsto\; \ket{j\,m} \otimes \ket{j\,m'}^{*},
$}
so that operators are represented as vectors in the tensor-product space $\mathcal H_j \otimes \mathcal H_j^{*}$. Since tensor-product spaces are generally reducible, it is natural to instead decompose $\mathcal H_j \otimes \mathcal H_j^{*}$ into the irreps of $\mathcal B(\mathcal H_j)$. This decomposition introduces operator space quantum numbers and yields a symmetry-adapted basis particularly suited to describing dissipative dynamics.

For collective spins, this reducibility takes the explicit form
\begin{equation}
\mathcal{B}(\mathcal H_j)
\;\cong\;
\mathcal H_j \otimes \mathcal H_j^\ast
\;\cong\;
\bigoplus_{k=0}^{2j} \mathcal{B}_k ,
\label{eqn:irreps}
\end{equation}
where each $\mathcal{B}_k$ transforms irreducibly under $\mathrm{SU}(2)$. The allowed values of $k$ follow from the Clebsch--Gordan decomposition of two spin-$j$ representations, $
j \otimes j = \bigoplus_{k=0}^{2j} k$.
Physically, this tensor-product structure may be viewed as the combination of two independent spin spaces. In the present case, however, the second space corresponds to the dual Hilbert space $\mathcal H_j^\ast$, which has important consequences for perturbative calculations (see Sec.~\ref{sec:perturbative_approach}). This decomposition highlights a key structural feature of open quantum dynamics: operator space splits into symmetry-resolved sectors that are not visible in a state-based description.

The generators of collective rotations in operator space are the superoperators implementing infinitesimal conjugation,
\begin{equation}
\mathcal K_\alpha[A] \equiv [J_\alpha,A],
\qquad \alpha=x,y,z,
\label{eq:adjoint_generator_definition}
\end{equation}
commonly referred to as the adjoint generators (see Appendix~\ref{sec:adjoint generators}). These generate rotations while acting on operators rather than states. The adjoint generators obey the same $\mathfrak{su}(2)$ algebra as the collective spin operators,
\begin{equation}
[\mathcal K_\alpha,\mathcal K_\beta]
=
i\epsilon_{\alpha\beta\gamma}\mathcal K_\gamma,
\end{equation}
and define the Casimir superoperator
\begin{equation}
\mathcal K^2 = \sum_{\alpha=x,y,z}\mathcal K_\alpha^2 .
\end{equation}
Within each irreducible sector $\mathcal B_k$, the Casimir superoperator acts diagonally with eigenvalue $k(k+1)$, providing the quantum number that labels the operator irreps.

The operator space analogues of angular momentum eigenstates are the spherical tensor operators $T^k_q$~\cite{Varshalovich1988, Blum2012, Hecht2000}. These form a basis of simultaneous eigenoperators of both $\mathcal K^2$ and $\mathcal K_z$,
\begin{equation}
\begin{split}
   & \mathcal{K}^2 [T^k_q] = \sum_{\alpha=x,y,z} [J_\alpha, [J_\alpha, T^k_q]]
   = k(k+1)\, T^k_q, \\
   & \mathcal{K}_z [T^k_q] = [J_z, T^k_q]
   = q\, T^k_q .
\end{split}
\end{equation}
The quantum numbers $k$ and $q$ correspond to the spherical tensor rank $k$ and its magnetic component $q$. For a fixed $k$, the allowed values of $q$ form a multiplet with $q=-k,-k+1,\ldots,+k$, giving $(2k+1)$ operators in each sector.

These quantum numbers admit an interpretation in terms of multipole moments, where $k$ denotes the order of the multipole \cite{Blum2012}. In this sense, $k$ labels the effective complexity of many-body operators. For example, the sector $k=1$ contains the dipole operators $J_x$, $J_y$, and $J_z$, which are captured by mean-field descriptions. Higher values of $k$ correspond to operators encoding progressively higher-order spin correlations. Expanding the density operator in the spherical tensor basis therefore amounts to decomposing it into contributions of increasing multipole order. In semiclassical approximations, this hierarchy is truncated by retaining only polynomial expectation values of $k=1$ operators (see Sec.~\ref{sec:perturbative_approach}).

Thus, this framework introduces a set of operator space quantum numbers that label operators independently of a state-based description. It also provides a symmetry-adapted operator basis that can be used for practical calculations. In the following section, we demonstrate the utility of this framework.

Throughout our work calligraphic symbols ($\mathcal A$) are used for superoperators acting on the Hilbert space of operators $\mathcal B (\mathcal H_j)$, whereas double-lined symbols ($\mathbb A$) are used for superoperators acting on the vectorized, Liouville space $\mathcal H_j \otimes \mathcal H_j^*$ \cite{Gyamfi2020}. In addition, the vectorized form of an operator $O$ is denoted via $\ket{O}\rangle$.

\section{Weak Symmetries and Operator Space Transport}
\label{sec:classification_of_collective_spin_dynamics}

\begin{figure*}[t]
\centering
\begin{tikzpicture}[
    x=1cm, y=1cm,
    font=\small,
    >=Latex,
    tensor/.style={
        draw,
        rounded corners=2pt,
        fill=blue!10,
        minimum width=1.0cm,
        minimum height=0.65cm,
        align=center,
        line width=0.6pt
    },
    groupbox/.style={
        draw,
        rounded corners=3pt,
        fill=gray!6,
        inner sep=4pt,
        line width=0.6pt
    },
    label/.style={font=\small},
    lattice/.style={
        circle,
        draw,
        fill=blue!12,
        minimum size=9mm,
        inner sep=0pt,
        line width=0.6pt
    },
    qhop/.style={
        -Latex,
        line width=0.8pt,
        opacity=0.45,
        color=orange!85!black
    },
    khop/.style={
        -Latex,
        line width=0.8pt,
        opacity=0.40,
        color=black!65
    },
    maparrow/.style={
        -Latex,
        line width=0.9pt,
        opacity=0.45,
        color=black!70
    }
]


\def\xplus{-7.8}
\def\xcenter{-4.5}
\def\xbrace{-6.8}

\node[font=\bfseries] at (\xplus -0.5, 3.5) {a)};
\node[font=\bfseries] at (\xplus + 10, 3.5) {b)};

\node[label] at (\xplus,2.7) {$\rho =$};

\node[groupbox] (sing) at (\xcenter,2.7) {
    \begin{tikzpicture}[x=1cm,y=1cm]
    \node[tensor] at (0,0) {$T^0_0$};
    \end{tikzpicture}
};
\node[label, anchor=west] at (\xcenter+1.2,2.7) {$k=0$};

\node[label] at (\xplus,1.5) {$+$};

\node[groupbox] (trip) at (\xcenter,1.5) {
    \begin{tikzpicture}[x=1cm,y=1cm]
    \node[tensor] at (-1.1,0) {$T^1_{-1}$};
    \node[tensor] at (0,0) {$T^1_0$};
    \node[tensor] at (1.1,0) {$T^1_{1}$};
    \end{tikzpicture}
};
\node[label, anchor=west] at (\xcenter+2.2,1.5) {$k=1$};

\node[label] at (\xplus,0.0) {$+$};

\node[groupbox] (quin) at (\xcenter,0.0) {
    \begin{tikzpicture}[x=1cm,y=1cm]
    \node[tensor] at (-2.2,0) {$T^2_{-2}$};
    \node[tensor] at (-1.1,0) {$T^2_{-1}$};
    \node[tensor] at (0,0) {$T^2_0$};
    \node[tensor] at (1.1,0) {$T^2_{1}$};
    \node[tensor] at (2.2,0) {$T^2_{2}$};
    \end{tikzpicture}
};
\node[label, anchor=west] at (\xcenter+3.3,0.0) {$k=2$};

\node[label] at (\xplus, -1.1) {$+$};
\node[label, anchor=west] at (\xcenter,-1.1) {$\vdots$};
\draw[maparrow] (-0.1,0.55) -- (1.55,0.55)
node[midway, above=2pt, font=\small] {mapped to};


\draw[->] (3.0,-0.6) -- (8.4,-0.6) node[right] {$q$};
\draw[->] (5.7,-0.9) -- (5.7, 2.8) node[above] {$k$};

\node[lattice] (t00) at (5.7,0.0) {$T^0_0$};

\node[lattice] (t1m1) at (4.6,1.0) {$T^1_{-1}$};
\node[lattice] (t10)  at (5.7,1.0) {$T^1_0$};
\node[lattice] (t11)  at (6.8,1.0) {$T^1_{1}$};

\node[lattice] (t2m2) at (3.5,2.0) {$T^2_{-2}$};
\node[lattice] (t2m1) at (4.6,2.0) {$T^2_{-1}$};
\node[lattice] (t20)  at (5.7,2.0) {$T^2_0$};
\node[lattice] (t21)  at (6.8,2.0) {$T^2_{1}$};
\node[lattice] (t22)  at (7.9,2.0) {$T^2_{2}$};

\draw[dashed, line width=0.5pt, opacity=0.4] (5.7,0.0) -- (7.9,2.0);
\draw[dashed, line width=0.5pt, opacity=0.4] (5.7,0.0) -- (3.5,2.0);


\draw[qhop] (t1m1.east) -- (t10.west);
\draw[qhop] (t10.west)  -- (t1m1.east);
\draw[qhop] (t10.east)  -- (t11.west);
\draw[qhop] (t11.west)  -- (t10.east);

\draw[qhop] (t2m2.east) -- (t2m1.west);
\draw[qhop] (t2m1.west) -- (t2m2.east);

\draw[qhop] (t2m1.east) -- (t20.west);
\draw[qhop] (t20.west)  -- (t2m1.east);

\draw[qhop] (t20.east)  -- (t21.west);
\draw[qhop] (t21.west)  -- (t20.east);

\draw[qhop] (t21.east)  -- (t22.west);
\draw[qhop] (t22.west)  -- (t21.east);


\draw[khop] (t00.north) -- (t10.south);
\draw[khop] (t10.south) -- (t00.north);

\draw[khop] (t10.north) -- (t20.south);
\draw[khop] (t20.south) -- (t10.north);

\draw[khop] (t1m1.north) -- (t2m1.south);
\draw[khop] (t2m1.south) -- (t1m1.north);

\draw[khop] (t11.north) -- (t21.south);
\draw[khop] (t21.south) -- (t11.north);

\end{tikzpicture}
\caption{
Schematic Representation of the Spherical Tensor Decomposition of the Density Operator. 
a) The density operator expanded into irreducible tensor multiplets of rank $k$, where each multiplet contains $2k+1$ components labeled by $q=-k,\dots,+k$. 
b) The same decomposition may be viewed as a triangular lattice in operator space, where each node corresponds to a spherical tensor mode $T^k_q$. Orange horizontal arrows indicate intra-rank ($q$) mixing, while grey arrows indicate inter-rank ($k$) mixing.
}
\label{fig:spherical_tensor_lattice}
\end{figure*}
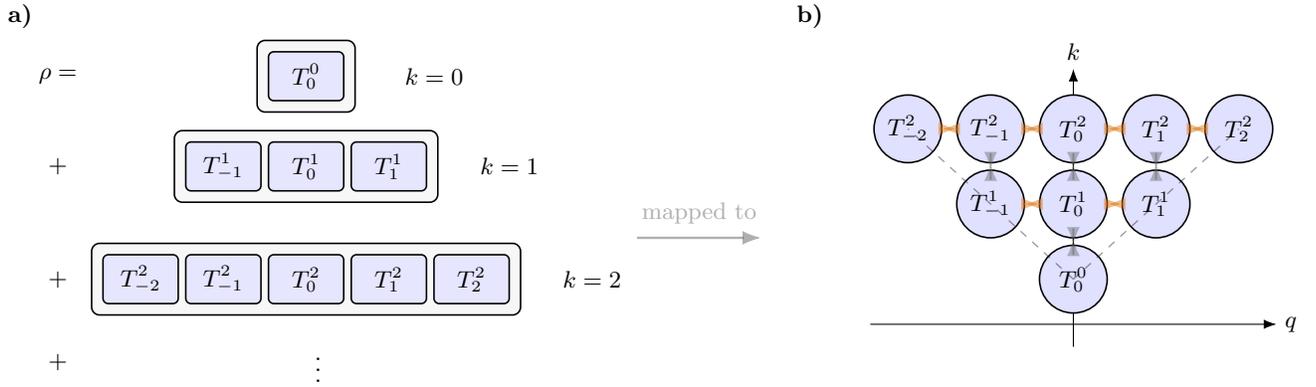
In this section, we show how the adjoint generators introduced in
Sec.~\ref{sec:representation_theory_for_collective_spins}, together
with their Casimir superoperator, can define weak symmetries of the
Liouvillian~\cite{Buca2012, Albert2014}. Commutation with $\mathcal K^2$ or $\mathcal K_z$
conserves the tensor rank $k$ or component $q$, respectively, thereby
imposing selection rules on transport in operator space. These weak
symmetries therefore provide a simple way to distinguish
collective spin models according to their allowed transport
structures. We first present a general overview of this framework and
then apply it to two explicit examples, demonstrating how the
preservation or breaking of these weak symmetries modifies the
connectivity of an ``operator lattice''. We emphasize that these symmetry constraints determine which transport processes are allowed, but do not by themselves determine the realized dynamical regime, which also depends on the Liouvillian matrix elements, model parameters, and spectral properties.

We analyze these systems by expanding $\rho$ in the spherical tensor basis $\{T^k_q\}$ as,
\begin{equation}
    \rho(t) = \sum_{k=0}^{2j} \sum_{q=-k}^{+k} a_{k,q} (t) \, T^k_q,
    \label{eq:spherical_tensor_basis}
\end{equation}
as shown in Fig.~\ref{fig:spherical_tensor_lattice}~\hyperref[fig:spherical_tensor_lattice]{a)}.

The master equation $\tfrac{d}{dt}\rho = \mathcal L \rho$ can be expressed in terms of the expansion coefficients $\{a_{k,q}\}$ as
\begin{equation}
    \frac{d}{dt} a_{k,q}(t)
    =
    \sum_{k'=0}^{2j} \sum_{q'=-k'}^{+k'}
    \mathcal L_{(k,q)(k',q')} \, a_{k',q'}(t),
    \label{eq:liouvillian_spherical_tensor_basis_general}
\end{equation}
with matrix elements
\begin{equation}
\mathcal L_{(k,q)(k',q')} \equiv
\mathrm{Tr}\!\left[(T^k_q)^\dagger \, \mathcal L \, T^{k'}_{q'}\right].
\label{eq:liouvillian_kq_matrix_elements}
\end{equation}
Introducing a composite index $\alpha \equiv (k,q)$, this can be written compactly as
\begin{equation}
\frac{d}{dt} a_\alpha(t) = \sum_\beta \mathcal L_{\alpha\beta} \, a_\beta(t),
\end{equation}
which has the form of a non-Hermitian tight-binding model. In this representation, the coefficients $a_\alpha$ play the role of amplitudes on an emergent operator space lattice, whose sites are labeled by the spherical tensor modes. The matrix elements $\mathcal L_{\alpha\beta}$ therefore act as effective hopping amplitudes from site $\beta$ to site $\alpha$ on the lattice. As we show below, the structure of the Liouvillian leads to short-range couplings in $(k,q)$, giving rise to a local hopping model on the operator lattice [see Fig.~\ref{fig:spherical_tensor_lattice}\hyperref[fig:spherical_tensor_lattice]{b)}]. For a derivation of Eq.~(\ref{eq:liouvillian_spherical_tensor_basis_general}) see Appendix~\ref{app:general_evol_eqn_in_tensor_basis}.

We now illustrate the utility of this framework using two explicit models that share the same coherent drive but differ in their choice of collective jump operator. This difference changes the weak symmetry structure of the Liouvillian and hence the allowed transport channels on the $(k,q)$ lattice.

\subsection{Driven Collective Spins with Dephasing: Collective Precession}
\label{sec:collective_precession_classification}

We first consider the case in which tensor rank $k$ is conserved, while the component $q$ is not,
\begin{equation}
[\mathcal L,\mathcal K^2]=0,
\qquad
[\mathcal L,\mathcal K_z]\neq0.
\end{equation}
The first condition makes each fixed-$k$ subspace invariant under the
Liouvillian evolution, whereas the second permits off-diagonal matrix elements that couple different $q$ components within these subspaces. Consequently, operator weight may be redistributed within each multiplet but cannot be transported between different tensor ranks.

Therefore, sectors with different $k$ evolve independently, and Eq.~(\ref{eq:liouvillian_spherical_tensor_basis_general}) reduces to a block-diagonal form,
\begin{equation}
    \frac{d}{dt} \mathbf{a}^{(k)} (t) = \mathcal L^{(k)} \, \mathbf{a}^{(k)} (t),
\end{equation}
where $\mathbf a^{(k)}=(a_{k,-k},\ldots,a_{k,k})^T$, and $\mathcal L^{(k)}$ denotes the restriction of $\mathcal L_{(k,q)(k',q')}$ to the subspace with fixed $k$, i.e., the $(2k+1)$-dimensional block with matrix elements $\bigl(\mathcal L^{(k)}\bigr)_{q q'} = \mathcal L_{(k,q)(k,q')}$.

A minimal realisation of coherent $q$-mixing at fixed $k$ is provided by a transverse collective field, $H=\Omega J_x$, with a jump operator along the quantization axis,
\begin{equation}
\mathcal L[\rho]
=
-i\Omega[J_x,\rho]
+
\frac{\Gamma}{2j}\!\left(
J_z \rho J_z
-\tfrac12\{J_z^2,\rho\}
\right).
\label{eq:L_Jx_Jz}
\end{equation}
The Hamiltonian term generates rotations about the $x$ axis and explicitly breaks conservation of $q$ with respect to $J_z$. As a result, $k$ remains a good quantum number, but $q$ is dynamically mixed within each fixed-$k$ sector. In the $\{T^k_q\}$ basis, the dynamics maps onto a local, non-Hermitian hopping model,
\begin{equation}
    \frac{d}{dt} a_{k,q}
    = -i \Omega \, \mathcal W_{k,q}
      - \frac{\Gamma}{4j} \, q^2 \, a_{k,q},
\label{eq:collective_precession_hopping_model}
\end{equation}
with
\begin{equation}
\mathcal W_{k,q}
=
w_+(k,q-1)\,a_{k,q-1}
+
w_-(k,q+1)\,a_{k,q+1}.
\end{equation}
Here, $w_{\pm}(k,q)
\equiv
\tfrac{1}{2}\sqrt{k(k+1) - q(q\pm1)}$
are position-dependent couplings generated by $\mathcal K_x$. The terms in $\mathcal W_{k,q}$ admit a hopping interpretation: the first term describes transport from $(k,q-1)$ to $(k,q)$ with amplitude $a_{k,q-1}$ and coupling strength $w_+(k,q-1)$, while the second term describes transport from $(k,q+1)$ to $(k,q)$ with amplitude $a_{k,q+1}$ and coupling $w_-(k,q+1)$. These couplings are reciprocal in $q$, satisfying $w_+(k,q)=w_-(k,q+1)$. Consequently, within each fixed-$k$ sector the dynamics corresponds to nearest-neighbor hopping along the $q$-direction, with $q$-dependent on-site contributions from the Lindbladian dissipator $\mathcal D$. For a detailed derivation of this mapping see Appendix~\ref{app:collective_precession_mapping}.

The operators $J_x$, $J_y$, and $J_z$ form the $k=1$ operator multiplet, so their dynamics is entirely determined by the
$q\in\{-1,0,+1\}$ components of this sector. The corresponding
coefficients $(a_{1,-1},a_{1,0},a_{1,1})$ obey
\begin{equation}
    \begin{split}
        \frac{d}{dt}a_{1,0} (t) & = -\frac{i \Omega}{\sqrt{2}} (a_{1,1}(t)  + a_{1,-1}(t) ), \\
        \frac{d}{dt}a_{1,1} (t) & = -\frac{i \Omega}{\sqrt{2}} a_{1,0}(t)  - \frac{\Gamma}{4j} a_{1,1}(t) , \\
        \frac{d}{dt}a_{1,-1} (t) & = -\frac{i \Omega}{\sqrt{2}}  a_{1,0}(t)  - \frac{\Gamma}{4j} a_{1,-1}(t) . \\
    \end{split}
    \label{eq:odes_for_exp_coeffs_q_mixing}
\end{equation}
For a complete discussion of the eigenvalues and eigenvectors of this system see Appendix~\ref{app:q_mixing_eigensystem}. When viewed in terms the observables $J_x, J_y$ and $J_z$, this results in a dynamics where $J_x$ is decoupled and exhibits exponential decay. $(J_y, J_z)$ are coupled together and exhibit decaying oscillations with frequency $\Omega \sqrt{1-(\kappa/2)^2}$, where $\kappa \equiv \Gamma/(4j\Omega)$. The decay factor is given by $\Gamma/(8j)$, which vanishes in the thermodynamic limit ($j\to \infty$). In this limit, both $J_y$ and $J_z$ exhibit persistent oscillations, leading to collective precession. An exceptional point emerges at $\kappa=2$, where oscillatory eigenmodes coalesce and the dynamics becomes pure relaxation~\cite{Nakanishi2023, Nemethb2025}. This example explicitly demonstrates that the weak symmetry structure fixes the allowed transport channels but does not by itself determine whether the resulting dynamics is oscillatory or relaxational.

The main consequence of the weak symmetry generated by $\mathcal K^2$ is that operator weight cannot be transferred between different tensor ranks. The initial distribution across $k$ sectors therefore cannot be dynamically redistributed, and observables in the $k=1$ multiplet cannot acquire contributions from higher-rank modes. This restricted connectivity is the transport property predicted by the symmetry analysis; the frequencies, decay rates, and exceptional points within each block must instead be determined from the corresponding matrix elements of $\mathcal L^{(k)}$.

In the following subsection, we consider a model for which
$[\mathcal L,\mathcal K^2]\neq0$. Breaking this weak symmetry permits inter-rank transport and converts the independent one-dimensional $q$ chains into a coupled two-dimensional operator space lattice.
Whether this enlarged connectivity produces BTC or overdamped behavior remains a dynamical question governed by the model parameters and Liouvillian spectrum.

\subsection{Driven Collective Spins with Decay: BTC Model}
\label{sec:btc_classification}
So far, we have considered a Liouvillian that commutes with $\mathcal K^2$, such that each tensor-rank sector evolves independently. We now turn to a model in which $\mathcal K^2$ is not conserved. Breaking this weak symmetry removes the fixed-$k$ block structure and permits transport between different tensor ranks.

The canonical form of the BTC model is given by~\cite{Iemini2018}
\begin{equation}
    \mathcal L[\rho]
    =
    -i\Omega[J_x,\rho]
    +
    \frac{\Gamma}{2j}\!\left(
    J_- \rho J_+
    -\tfrac12\{J_+J_-,\rho\}
    \right),
\label{eq:canonical_BTC}
\end{equation}
where the jump operator $J_-$ describes collective decay. In contrast to the $J_z$ dissipator considered above, its action does not preserve the adjoint Casimir superoperator,
\begin{equation}
    [\mathcal D,\mathcal K^2]\neq0.
\end{equation}
Consequently, for $\Gamma\neq0$,
\begin{equation}
    [\mathcal L,\mathcal K^2]\neq0.
\end{equation}
The coherent drive generated by $J_x$ also breaks conservation of the
tensor component with respect to the $z$ axis. Thus, for
$\Gamma,\Omega\neq0$,
\begin{equation}
    [\mathcal L,\mathcal K^2]\neq0,
    \qquad
    [\mathcal L,\mathcal K_z]\neq0.
\label{eq:BTC_weak_symmetries}
\end{equation}
Therefore, neither $k$ nor $q$ is conserved, and transport along both directions of the operator space lattice is allowed. In the limiting case $\Omega=0$, conservation of $q$ is restored and the dynamics reduces to dissipative transport at fixed $q$.

We emphasize that Eq.~(\ref{eq:BTC_weak_symmetries}) establishes only
that $\mathcal K^2$ and $\mathcal K_z$ are not weak symmetries of this Liouvillian; it does not exclude the existence of other weak or
strong symmetries. For example, the total spin operator $J^2$ remains a strong symmetry because it commutes separately with the Hamiltonian and the jump operator. Since we work within a fixed-$j$ sector this strong symmetry does not constrain transport between operator sectors labelled by $k$ and $q$.

 The action of $\mathcal D$ couples neighboring tensor-rank sectors, thereby permitting inter-rank transport. Together with the $q$-mixing generated by the coherent drive, this produces a
two-dimensional connectivity structure on the $(k,q)$ lattice.
Specifically, $\mathcal D$ generates nearest-neighbor hopping along
the $k$ direction together with local on-site contributions
(see Appendix~\ref{app:dissipator_induced_k_hopping}). Note that nearest-neighbor hopping in $k$ is a consequence of linear jump operators. More general jumps, such as $J_+^2$, may generate longer-range couplings in $k$, with the allowed hopping range determined by the irreducible tensor addition rules discussed in Appendix~\ref{app:dissipator_induced_k_hopping}.

\begin{figure}[t!]
    \centering
    \includegraphics[width=\linewidth]{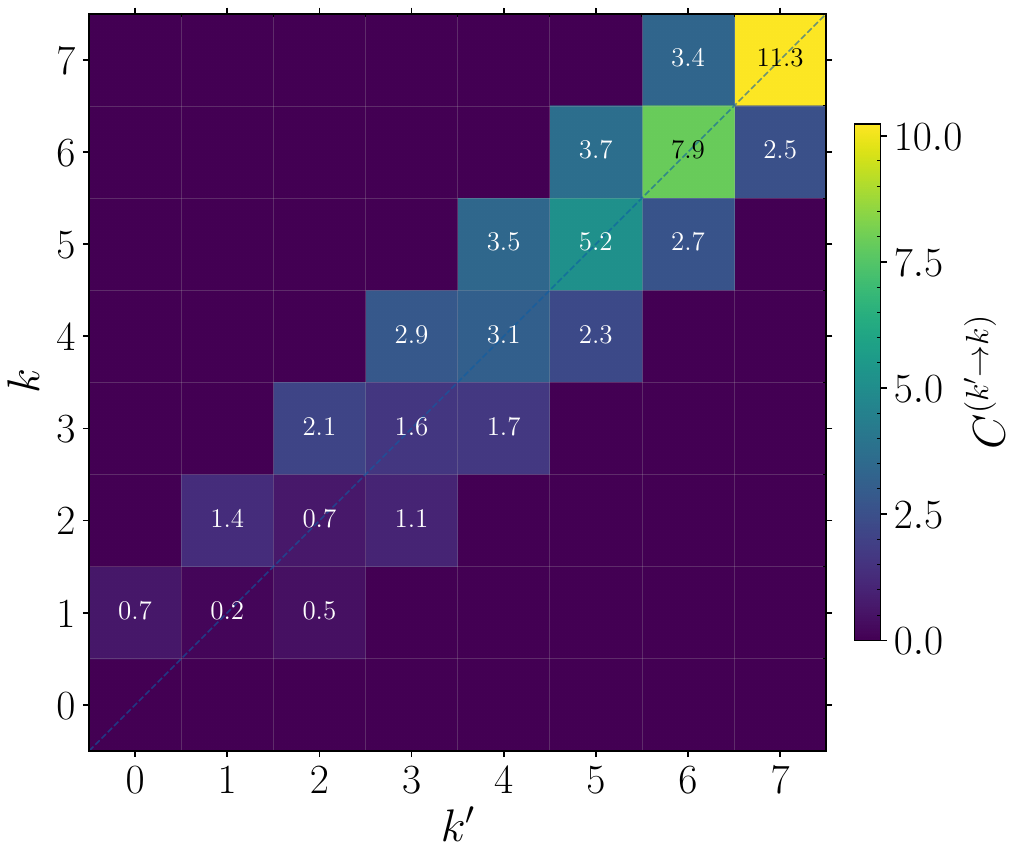}
    \caption{Rank Coupling Matrix. A numerical visualization of the coupling strengths between irreducible sectors $k'\to k$, quantified by $C^{(k' \to k)}$~[Eq.~(\ref{eq:rank_coupling_strengths})] and represented by the color scale. Results are shown for total spin $j=7/2$ ($k=0,1,\ldots,7$) with $\Gamma = \Omega=1$. Numerical values are displayed for non-zero entries to highlight the allowed transitions between sectors.}
    \label{fig:rank_coupling_matrix}
\end{figure}

The Lindbladian dynamics can therefore be represented as a nearest-neighbour non-Hermitian hopping problem, where $k$ and $q$ act as effective coordinates on an emergent two-dimensional operator lattice. In this representation, Eq.~(\ref{eq:canonical_BTC}) maps to
\begin{equation}
\frac{d}{dt} a_{k,q} = -i\Omega\,\mathcal W_{k,q} + \frac{\Gamma}{2j}\,\mathcal T_{k,q},
\label{eq:non_hermitian_hopping_mapping}
\end{equation}
with
\begin{equation}
\begin{split}
\mathcal T_{k,q}
= \;
& t_+(k-1,q)\,a_{k-1,q}
+
t_-(k+1,q)\,a_{k+1,q} \\
& -
\gamma(k,q)\,a_{k,q}.
\label{eq:t+t-_and_gamma}
\end{split}
\end{equation}
Here, $\gamma(k,q)$ denotes an on-site term at the coordinate $(k,q)$, while $t_{\pm}(k,q)$ describe hopping between $(k,q)$ and $(k\pm1,q)$. In the present case, hopping along the $k$-direction is non-reciprocal, $t_+(k,q)\neq t_-(k+1,q)$ (see Fig.~\ref{fig:rank_coupling_matrix}), giving rise to an effective non-Hermitian transport problem in $k$. The hopping amplitudes $t_\pm(k,q)$ generally depend on the total spin $j$. Since our analysis is performed within a fixed-$j$ sector, for qualitative discussions this may be suppressed for notational simplicity. The full expressions $t_\pm(k,q,j)$ together with their derivation, are presented in Appendix~\ref{app:dissipator_induced_k_hopping}. On the other hand, $\gamma(k,q)$ is independent of the total spin $j$, yielding
\begin{equation}
\gamma(k,q)=\frac{1}{2}\Bigl[k(k+1)-q^2\Bigr].
\label{eq:onsite_decay_explicit_form}
\end{equation}
This corresponds to an inhomogeneous on-site decay in operator space: the damping increases with rank $k$, while for fixed $k$ it is largest at the centre of a multiplet ($q=0$) and weakest at its boundaries ($q=\pm k$).

In contrast, the coherent contribution remains identical to the collective precession case, with $w_\pm(k,q)$ generating reciprocal hopping along the $q$-direction between $(k,q)$ and $(k,q\pm1)$. The corresponding hopping amplitudes can also be extracted numerically from the Liouvillian matrix elements in the $\{T^k_q\}$ basis, as shown in Fig.~\ref{fig:rank_coupling_matrix}.

A numerical visualization of the coupling strengths between different ranks $k'$ and $k$, $C^{(k'\to k)}$, is shown in Fig.~\ref{fig:rank_coupling_matrix}. To construct $C^{(k' \to k)}$, we first define the projectors onto the eigenspace of $\mathcal K^2$ with eigenvalues $k(k+1)$ as $\mathbb P^{(k)}$. Then, by introducing 
\begin{equation}
   \mathbb D^{(k'\to k)}  = \mathbb P^{(k)} \, \mathbb D \, \mathbb P^{(k')}
\end{equation}
as the flow of operator weight from the sector $k'$ to $k$ generated by the Lindbladian dissipator $\mathbb D$ (acting on the vectorized $\ket{\rho\rangle}$), we define $C^{(k' \to k)}$ as
\begin{equation}
    C^{(k' \to k)}= \sqrt{\sum_{qq'} \abs{(\mathbb D^{(k' \to k)})_{qq'} }^2}
    \label{eq:rank_coupling_strengths}
\end{equation}
such that we sum over all possible $q$-channels to go from $k'$ to $k$. Therefore, this represents a measure of the overall coupling strength between the two sectors. For further details on the construction of the projectors, see Appendix~\ref{sec:projectors}.

In Fig.~\ref{fig:rank_coupling_matrix}, the first off-diagonal elements encode effective, nearest-neighbor couplings between sectors. These couplings are asymmetric, which is due to the non-reciprocal hopping amplitudes, $t_+(k,q) \neq t_-(k+1,q)$. The diagonal elements correspond to dissipative contributions that preserve $k$, captured by the on-site decay rates $\gamma(k,q)$ [c.f. Eq.(\ref{eq:onsite_decay_explicit_form})].

Therefore, we have demonstrated how the dynamics of the canonical BTC model [Eq.~(\ref{eq:canonical_BTC})] can be formulated
as a nearest-neighbor transport problem on the two-dimensional $(k,q)$ lattice. The symmetry analysis predicts that operator weight is no longer confined to fixed-rank multiplets and that low-rank observables may hybridize with higher-rank operator sectors, see Sec.~\ref{sec:dissipation_driven_hybridization} for more detail on the hybridization. These symmetries therefore constrain the allowed forms of collective spin dynamics, whose specifics are determined by the ratio $\Gamma / \Omega$. We expand on this in Sec.~\ref{sec:perturbative_approach} by discussing how our approach can be used to study the extreme time crystal regime, which is characterized by $\Gamma \ll \Omega$.

\section{Initial State Independent Dynamics: BTC Model}
\label{sec:initial_state_independent_dynamics}
The example considered in Sec.~\ref{sec:collective_precession_classification} illustrates how oscillatory dynamics can arise in the strong-driving regime, \(\Gamma\ll\Omega\) . The coherent term generated by \(J_x\) produces imaginary couplings between different \(q\) components within each fixed-\(k\) multiplet. For \(k=1\), these couplings give the familiar collective precession of \(J_y\) and \(J_z\) about the \(x\)-axis. The same mechanism applies more generally to every tensor-rank sector. Higher-rank multiplets, however, contain several distinct non-zero values of \(\vert q\vert\), and can therefore support multiple oscillation frequencies, beating, and higher harmonics; see Ref.~\cite{nemeth2026topologicalboundarytimecrystal} for further discussion.

When $k$ is conserved, these oscillations are necessarily initial state dependent. Each multiplet evolves independently according to $
    \mathbf a^{(k)}(t)
    =
    \bigl(a_{k,-k}(t),\ldots,a_{k,k}(t)\bigr)^T,
$
so a sector can evolve only if it is populated initially. For example, since the $k=1$ multiplet is spanned by $J_x,J_y$ and $J_z$, an initial state satisfying
\begin{equation}
    \Tr\!\left(\rho(0)J_\alpha\right)=0,
    \qquad \alpha=x,y,z,
\end{equation}
has no initial weight in this sector. If $[\mathcal L,\mathcal K^2]=0$, no other rank sector can subsequently populate it, and the $k=1$ contribution therefore remains zero.

The situation changes qualitatively once the weak symmetry generated by $\mathcal K^2$ is broken. When
$
    [\mathcal L,\mathcal K^2]\neq0,
$
operator weight can be transported between different tensor ranks. For the BTC model, the Lindbladian dissipator with a linear collective jump operator produces nearest-neighbour hopping in $k$, so the dynamics of a given multiplet is no longer determined solely by its initial amplitudes: it can also be fed dynamically by neighbouring sectors.

Crucially, such feeding contains both initial state dependent and initial state independent contributions. To distinguish them, we decompose the density operator into its identity and traceless parts~\cite{Byrd2003, Kimura2003, Bertlmann2008,Byrd2011},
\begin{equation}
    \rho(t)=\frac{\mathbf 1}{d}+\delta\rho(t),
\end{equation}
where \(d=2j+1\) and
\begin{equation}
    \delta\rho(t)
    =
    \sum_{k\ge1}\sum_{q=-k}^{k}
    a_{k,q}(t)T_q^k.
\end{equation}
The coefficient of the identity is fixed by normalization and is therefore the same for every physical initial state.

Trace preservation implies $
    \Tr\!\left(\mathcal L[\rho]\right)=0
$~\cite{Rivas_2012}. In particular, if $\delta \rho$ is traceless, then $\mathcal{L}[\delta \rho]$ is also traceless, so the traceless operator subspace is invariant under the evolution. The identity, however, need not itself be invariant: for a non-unital Lindbladian,
$
    \mathcal L[\mathbf 1]\neq0.
$
Because $\mathcal L[\mathbf 1]$ is necessarily traceless, the identity sector may act as a source for the traceless subspace, whereas trace preservation forbids the reverse process.

In the decomposition into identity and traceless sectors, the Liouvillian therefore takes the block-triangular form~\cite{Bittner2026}
\begin{equation}
    \mathcal L
    =
    \begin{bmatrix}
        0 & 0\\
        \mathcal S & \mathcal M
    \end{bmatrix},
\end{equation}
where $\mathcal M$ governs evolution entirely within the traceless subspace, while
$
    \mathcal S[\mathbf 1]=\mathcal L[\mathbf 1]
$
encodes the non-unital source.

The traceless component consequently obeys the inhomogeneous equation~\cite{Schirmer2010, Schirmer2012, Yuan2012}
\begin{equation}
    \frac{d}{dt}\delta\rho(t)
    =
    \mathcal M\,\delta\rho(t)
    +\frac{1}{d}\mathcal S[\mathbf 1].
\end{equation}
Its formal solution is
\begin{equation}
    \delta\rho(t)
    =
    e^{\mathcal Mt}\delta\rho(0)
    +
    \frac{1}{d}
    \int_0^t ds\,
    e^{\mathcal M(t-s)}
    \mathcal S[\mathbf 1].
\end{equation}
This expression makes the distinction transparent. The first term is the homogeneous contribution and retains information about the initial traceless components of the density operator. The second is an inhomogeneous contribution generated entirely by the non-unital source. Since the identity component \(\mathbf 1/d\) is fixed by normalization, this contribution is common to all physical initial states.

For the BTC model, Appendix~\ref{app:dissipator_induced_k_hopping} gives
\begin{equation}
    \mathcal S[\mathbf 1]
    =
    -\frac{\Gamma}{j}J_z.
\end{equation}
Thus, collective decay continuously injects operator weight from the $k=0$ identity sector into the $k=1$, $q=0$ component, since $T_0^1\propto J_z$. The coherent $J_x$ drive then mixes this component with $q=\pm1$ within the same multiplet, generating oscillatory $J_y$ and $J_z$ dynamics. Importantly, this mechanism does not require any initial population of the $k=1$ sector.

The same sourced weight can subsequently propagate to higher tensor ranks through dissipative hopping in $k$. Consequently, higher-rank oscillatory components can also be generated dynamically even when they are absent from the initial state. The maximally mixed state,
$
    \rho(0)=\mathbf 1 / d,
$
provides the clearest example: initially, every traceless amplitude $a_{k,q}(0)$ vanishes, yet non-unital dissipation first generates the $k=1$ component and the combined $k$- and $q$-transport can then spread this weight throughout operator space~\cite{nemeth2026topologicalboundarytimecrystal}.

Non-unitality therefore provides a microscopic mechanism by which oscillatory modes can be populated independently of the initial traceless state $\delta \rho(0)$. In the operator space picture, the identity sector acts as a universal source, while coherent transport in $q$ can convert the sourced weight into oscillatory dynamics and dissipative transport in $k$ distributes it between different multipole sectors. Further details of these source-driven contributions and their dynamical behaviour are given in Ref.~\cite{nemeth2026topologicalboundarytimecrystal}.

Finally, we emphasize that this is a purely dissipative effect and cannot arise from unitary Hamiltonian dynamics. Indeed, for any Hermitian Hamiltonian $H$, one has $[H,\mathbf{1}] = 0$, so the identity sector remains invariant under coherent evolution. Consequently, the non-reciprocal transport discussed here originates entirely from the dissipative (Lindbladian) contribution.

\section{Relation to Semiclassical Approaches: BTC Model}
\label{sec:relation_to_semiclassical_approaches}
The time crystal phase of the canonical BTC model [Eq.~(\ref{eq:canonical_BTC})] was originally identified through the emergence of stable limit-cycle dynamics~\cite{Iemini2018,Piccito2021}. These approaches begin from the Heisenberg equations of motion for the collective spin operators $J_\alpha$. Since the resulting equations couple first moments to higher-order spin correlations, an approximation is required to obtain a closed set of equations.

The simplest closure is provided by the mean-field approximation~\cite{Gutzwiller1963, Krauth1992, Fazio2025},
\begin{equation}
    \langle J_\alpha J_\beta\rangle
    \approx
    \langle J_\alpha\rangle
    \langle J_\beta\rangle,
\end{equation}
where $\alpha,\beta=x,y,z$. Introducing the normalized magnetizations
\begin{equation}
    m_\alpha=\frac{\langle J_\alpha\rangle}{j},
\end{equation}
this factorization closes the hierarchy at the level of first moments and yields the nonlinear equations~\cite{Iemini2018,Piccito2021,nemeth2026topologicalboundarytimecrystal}
\begin{equation}
\begin{split}
    \frac{d}{dt}m_x &= \kappa m_xm_z,\\
    \frac{d}{dt}m_y &= -\Omega m_z+\kappa m_ym_z,\\
    \frac{d}{dt}m_z &= \Omega m_y-\kappa(m_x^2+m_y^2),
\end{split}
\end{equation}
with $\kappa=\Gamma/2$. In the thermodynamic limit, this approximation becomes exact for the class of initially uncorrelated states~\cite{Mori2013,Carollo2021, Fiorelli2023}. It therefore provides a particularly useful description of the symmetry-breaking and limit-cycle structure of the BTC phase. Note that any non-zero initial magnetization vector $\boldsymbol{m}(0)=(m_x(0),m_y(0),m_z(0))^T$ can, in principle, develop oscillatory dynamics through the nonlinear coupling between its components. For example, for an initial state polarized along the $x$ direction,
\begin{equation}
    \dot m_z(0)=-\kappa m_x^2(0)\neq 0,
\end{equation}
so a non-zero $m_z$ component can be generated dynamically even when $m_y(0)=m_z(0)=0$. This differs from collective precession [c.f. Eq.~(\ref{eq:L_Jx_Jz})], where $m_x$ decouples from $m_y$ and $m_z$, and an initially $x$-polarized state remains stationary in the thermodynamic limit (see Appendix~\ref{app:q_mixing_eigensystem} for more detail).

Our operator space representation provides a complementary way to understand what information is retained by this approximation. The first moments $m_{x,y,z}$ are completely determined by the $k=1$ spherical tensor multiplet,
$
    \{a_{1,-1},a_{1,0},a_{1,+1}\}.
$
By contrast, the sectors $k\ge2$ contain higher-order multipole moments and correlations. Mean-field factorization does not retain these higher-rank amplitudes as independent dynamical variables. Instead, their influence on the equations of motion is approximated by products of the $k=1$ expectation values.

This also clarifies its dependence on the initial state. Two states with identical first moments but different higher-order correlations are indistinguishable within a mean-field description, despite occupying different regions of the full operator space. In the particularly important case
$
    m_x(0)=m_y(0)=m_z(0)=0,
$
the mean-field variables contain no information about higher-rank components of the initial density operator. Consequently, dynamics originating from these correlations, or from their subsequent transport into the $k=1$ sector, cannot in general be reconstructed from the first-moment equations alone. This provides an operator space interpretation of the initial state restrictions identified in Ref.~\cite{Mukherjee2024}. Namely, initial states satisfying
\begin{equation}
    m_x(0)=m_y(0)=m_z(0)=0,
\end{equation}
equivalently $a_{1,q}(0)=0$ for $q=0,\pm1$, have no initial support in the $k=1$ sector.

A systematic improvement is obtained by extending the closure to second order. As demonstrated in Ref.~\cite{Mukherjee2024}, retaining second-order cumulants allows BTC oscillations to be recovered for a broader class of initial states~\cite{Kubo1962, Henschel2010, Kirton2018, Fazio2025}. In our accompanying work~\cite{nemeth2026topologicalboundarytimecrystal}, we show explicitly how the non-unital sourcing of the $k=1$ multiplet appears through the second-order cumulants, and that it is absent from the first-order mean-field description.

The cumulant hierarchy may therefore be viewed as a sequence of progressively refined closures of the full operator space dynamics: mean-field theory retains only first moments, second-order cumulant expansions additionally capture quadrupolar ($k=2$) correlations, and higher-order closures progressively restore further multipole sectors. By contrast, the spherical tensor formulation employed here retains the complete hierarchy of $k$ sectors from the outset. This is particularly useful for obtaining analytical results at finite system size, where mean-field theory is not generally exact and becomes asymptotically controlled only in the thermodynamic limit, subject to assumptions on the class of initial states~\cite{Carollo2021, Fiorelli2023, Mukherjee2024}. Moreover, perturbative approaches formulated directly in the angular momentum basis can still require numerical diagonalization~\cite{Nakanishi2023}. In Sec.~\ref{sec:perturbative_approach}, we instead exploit the irreducible tensor structure to derive an effective Lindbladian that provides an analytical description of the extreme time crystal regime. Retaining the full operator space structure also makes the weak symmetries of $\mathcal L$ explicit and provides a framework for identifying properties that emerge upon approaching the thermodynamic limit (see Sec.~\ref{sec:thermodynamic_limit_btc_model} for more detail).

\section{Perturbative Approach in the Extreme Time Crystal Phase}
\label{sec:perturbative_approach}
The above framework can be extended to a perturbative description of the dynamics. In this section, we outline how the spherical tensor representation can be used to construct an effective Liouvillian valid deep in the time crystal regime. To this end, we perform a perturbative expansion, treating the dissipative contribution to first order in $\Gamma/j$.

Following the Choi--Jamiołkowski isomorphism~\cite{Choi1975, Jamiolkowski1972}, we vectorize the density operator as $\rho \mapsto \ket{\rho}\rangle$~\cite{Gyamfi2020}. Under this mapping, the adjoint generators transform as
\begin{equation}
    \mathcal{K}_\alpha \;\mapsto\; \mathbb{S}_\alpha \equiv J_\alpha \otimes \mathbf{1} - \mathbf{1} \otimes J_\alpha^T,
    \label{eq:adjoint_generator_superspin}
\end{equation}
where $\mathbb{S}_\alpha$ ($\alpha = x,y,z$) define the components of an effective ``superspin'' representation, as introduced in Ref.~\cite{Nemethb2025}.

In that work, the notion of a \emph{superspin} was introduced by observing that the tensor-product space $\mathcal H_j \otimes \mathcal H_j^*$ can be interpreted as describing two coupled subsystems. The second subsystem corresponds to the dual Hilbert space, such that spin operators act as $J_\alpha \otimes \mathbf{1}$ on the original space and $-\mathbf{1} \otimes J_\alpha^{T}$ on the dual space.

In the limit $\Gamma \to 0$, Eq.~(\ref{eq:canonical_BTC}) reduces to purely unitary evolution generated by these two contributions, which can be combined into a single effective degree of freedom. This emergent object is referred to as the superspin~\cite{Nemethb2025}.

For $\Gamma \neq 0$, the dissipator induces an effective coupling between these subsystems and renormalizes their energy levels. Nevertheless, as shown in Ref.~\cite{Nemethb2025}, the dynamics can still be captured by an effective description in terms of the superspin to first order in $\Gamma/j$,
\begin{equation}
    \mathbb L_{\mathrm{eff}} = -i \Omega \, \mathbb S_x - \frac{\Gamma}{8j}\Bigl[ \mathbb S_x^2 + \mathbb S^2\Bigr].
    \label{eq:effective_vectorized_liouvillian}
\end{equation}
This framework has since been used to derive precision bounds for frequency estimation in continuously monitored BTCs and to show that these systems can surpass the Heisenberg limit~\cite{Oconnor2026,Jirasek2025}.

Here we show that the superspin method emerges directly from the adjoint representation and working in the spherical tensor basis. The superspin components shown in Eq.~(\ref{eq:adjoint_generator_superspin}) are equivalent to the adjoint generators, expressed in a form that acts on vectorized Liouville space. Consequently, Eq.~(\ref{eq:effective_vectorized_liouvillian}) can be written as
\begin{equation}
    \mathcal L_{\mathrm{eff}}
    =
    -i \Omega \mathcal K_x
    - \frac{\Gamma}{8j}\Bigl[ \mathcal K_x^2 + \mathcal K^2\Bigr].
    \label{eq:btc_effective_model}
\end{equation}
Its eigenoperators are given by $T^k_{q_x}$ with eigenvalues
\begin{equation}
    \lambda_{k,q_x}
    =
    -i\Omega\, q_x
    -
    \frac{\Gamma}{8j}\bigl[q_x^2 + k(k+1)\bigr],
    \label{eq:first_order_perturbation_eigenvalues}
\end{equation}
where the quantization axis has been rotated along the $x$-axis. To first order in $\Gamma$, the eigenvalues therefore form multiplets labeled by $k=0,1,\ldots,2j$ and $q_x=-k,\ldots,+k$. This structure explains the eigenspectrum of $\mathcal L$ observed in numerics~\cite{Iemini2018,Nakanishi2023, Nemethb2025}. In addition, the superspin basis $\{\ket{s, \, s_x}\rangle\}$ is equivalent to the vectorized spherical tensor basis such that $T^k_{q_x} \mapsto \ket{k, \, q_x}\rangle$.

Therefore, in the extreme time crystal limit, the leading-order BTC dynamics [Eq.~(\ref{eq:btc_effective_model})] reduces to the same qualitative form as Eq.~(\ref{eq:L_Jx_Jz})~\cite{Nemethb2025}. The key distinction is that Eq.~(\ref{eq:btc_effective_model}) features anisotropic rather than isotropic dephasing, as discussed in detail in Ref.~\cite{Nemethb2025}.

Next, we demonstrate how dissipation modifies the dynamics beyond the extreme time crystal limit.

\begin{figure*}[ht!]
    \centering
    \includegraphics[width=\linewidth]{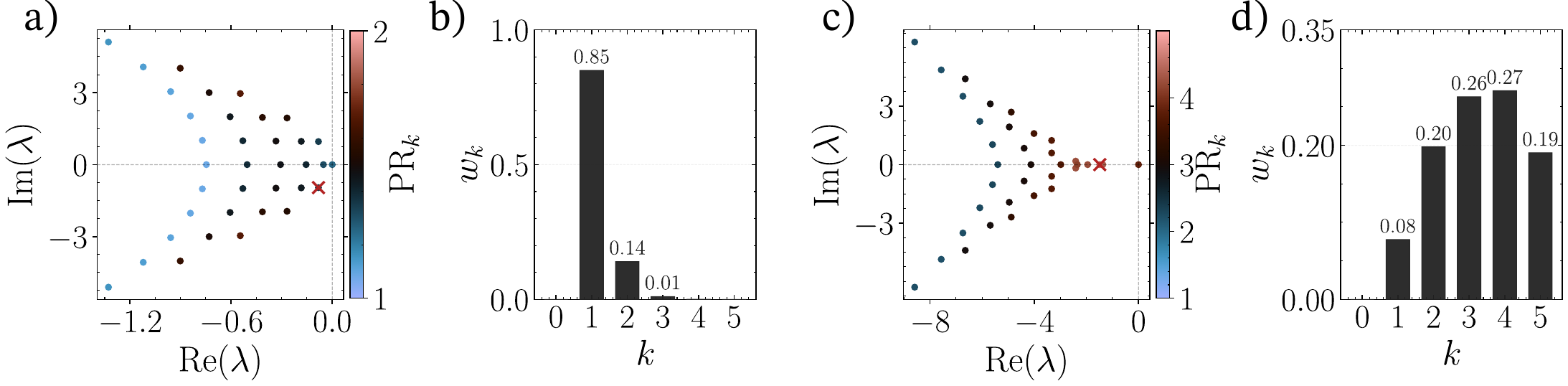}
    \caption{Hybridization of Spherical Tensor Modes. Results are shown for $j=5/2$ with $\Gamma / \Omega = 0.5$ (panels a, b) and $\Gamma / \Omega = 4$ (panels c, d). Panels (a, c) display the Liouvillian eigenvalues (filled circles) in the $(\mathrm{Re}(\lambda),\mathrm{Im}(\lambda))$ plane, where the color indicates the $k$-resolved participation ratio $\mathrm{PR}_k$ of each eigenmode. The eigenmode analyzed in panels (b, d) is marked by a red cross. Panels (b, d) show the corresponding distribution of weights across rank sectors, $w_k$, as a function of $k$. The bar heights and accompanying numerical labels indicate the total weight in each sector, with labels shown only for weights larger than $0.01$.}
    \label{fig:spherical_tensor_hybridization}
\end{figure*}

\section{Dissipation-Driven Hybridization in Operator Space}
\label{sec:dissipation_driven_hybridization}
In Sec.~\ref{sec:btc_classification} it was demonstrated how the BTC model can be mapped to a local hopping problem, which describes the transport of operator weight across a two-dimensional operator lattice. Eq.~(\ref{eq:non_hermitian_hopping_mapping}) highlights that the dissipator drives transport across $k$, therefore, in general, we expect the Liouvillian eigenmodes to be a superposition of many spherical tensor modes. We demonstrate this numerically in Fig.~\ref{fig:spherical_tensor_hybridization}.

We quantify how delocalized each eigenmode is in $k$ via the $k$-resolved participation ratio, $\mathrm{PR}^{(n)}_k$, which we define as
\begin{equation}
    \mathrm{PR}^{(n)}_k = \frac{1}{\sum_k \bigl(w^{(n)}_k\bigr)^2} , \qquad w^{(n)}_k = \sum_q \abs{c^{(n)}_{kq}}^2.
\end{equation}
This measures the effective number of rank sectors contributing to the $n^{\mathrm{th}}$ right eigenvector of $\mathbb L$. Here, $c_{kq} = \langle \langle k \, q \ket{r^{(n)}} \rangle$ is an expansion coefficient of this right eigenvector in the vectorized spherical tensor basis and $q$ is defined with respect to the quantization axis rotated along the $x$ direction, such that $q\equiv q_x$. If $\mathrm{PR}_k =1$, then only a single sector contributes and the eigenmode is localized. Alternatively, if an eigenmode is highly delocalized and has equal weights across all sectors then $\mathrm{PR}_k = 2j+1$.

\begin{figure*}[ht]
    \centering
    \includegraphics[width=\linewidth]{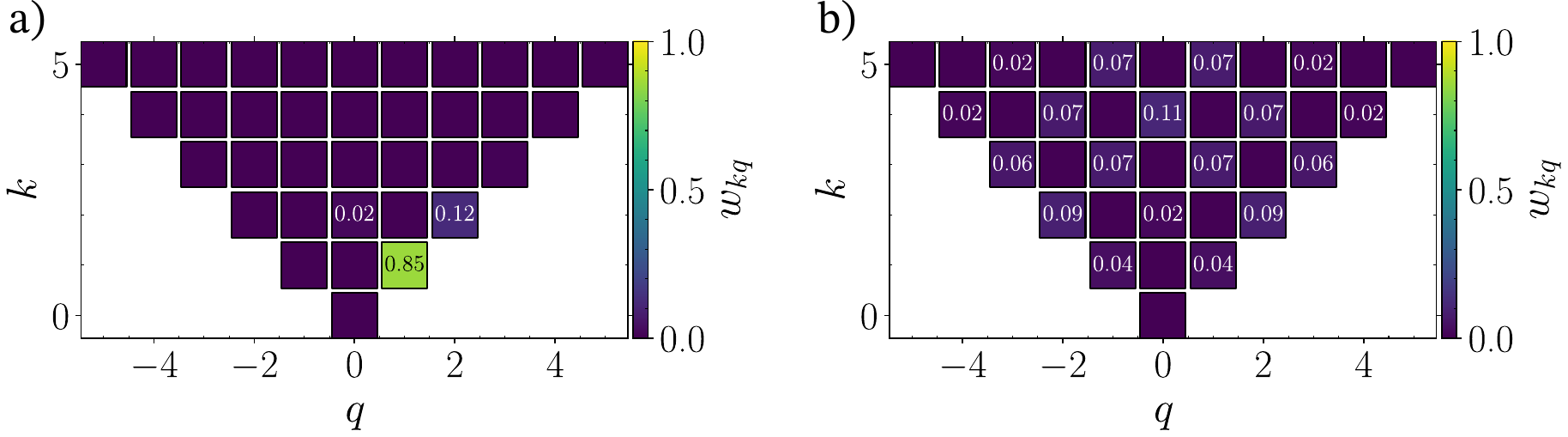}
    \caption{Heatmap on the $(k,q)$ Lattice. Results are shown for $j=5/2$ and $\Gamma / \Omega = 0.5$ (panel a) and $\Gamma / \Omega = 4$ (panel b). Weights $w_{kq}$ are shown for each site on the $(k,q)$ lattice, illustrating the spherical tensor modes contributing to a given eigenmode and their relative amplitudes. The eigenmode analyzed is marked by a red cross in Fig.~\ref{fig:spherical_tensor_hybridization} panels (a, c) respectively. Numerical labels are shown only for weights larger than $0.01$. The quantum number $q$ is defined with respect to a quantization axis rotated along the $x$ direction, such that $q \equiv q_x$.}
    \label{fig:kq_heatmap}
\end{figure*}

To further characterize this hybridization, we examine the distributions $w_k$ and $w_{kq}=|c_{kq}|^2$ for an eigenmode belonging to the slowest-decaying oscillatory pair. These are shown in Fig.~\ref{fig:spherical_tensor_hybridization} (panels b, d)  and Fig.~\ref{fig:kq_heatmap} respectively.

The dynamics of the model are governed by two competing mechanisms that dominate in opposite limits. The unitary evolution generated by the Hamiltonian localizes the dynamics in tensor rank $k$ while permitting transport in $q$. Consequently, the unperturbed eigenmodes are the vectorized spherical tensors $\ket{k\,q_x}\rangle$, and the dynamics correspond to collective precession within a fixed-$k$ multiplet. In contrast, the Lindblad dissipator induces non-reciprocal hopping between different tensor ranks, promoting transport and delocalization along the $k$-direction. 

The BTC phase emerges in the regime where transport along both directions is present. Here, dissipation is sufficiently strong to hybridize different $k$ sectors, yet weak enough that coherent transport in $q$ remains the dominant effect. As a result, operator weight spreads over a broad region of operator space while retaining the coherent dynamics necessary for persistent oscillations.

When $\Gamma \gg \Omega$, transport along $k$ dominates with coherent transport in $q$ becoming perturbatively suppressed. The multiplets therefore cease to undergo collective precession, leading to the disappearance of the BTC oscillations. This transport-based picture is fully consistent with finite-size spectral analysis: as the dissipation strength increases, complex conjugate eigenmode pairs [Fig.~\ref{fig:spherical_tensor_hybridization} panels a) and c)] merge at exceptional points, signalling the breakdown of oscillatory dynamics and the onset of an overdamped regime~\cite{Nakanishi2023, Nemethb2025}.

Furthermore, delocalization in $k$ hybridizes $\langle J_z(t)\rangle$ with higher-rank multiplets containing larger values of $q$. Since $q$ determines the oscillation frequencies of the Liouvillian eigenmodes [Eqs.~(\ref{eq:non_hermitian_hopping_mapping}) and (\ref{eq:first_order_perturbation_eigenvalues})], this hybridization causes $\langle J_z(t)\rangle$ to acquire frequency components originating from higher-order sectors, as observed in Ref.~(\cite{Iemini2018}). 

\section{Thermodynamic Limit of the BTC Model}
\label{sec:thermodynamic_limit_btc_model}

The $1/j$ Kac scaling of the Lindblad dissipator suggests that dissipative effects vanish in the thermodynamic limit, $j\rightarrow\infty$. However, the operator space representation reveals a more subtle behaviour: different contributions generated by the dissipator scale differently with $j$, such that local damping vanishes while dissipative hopping remains finite. As we highlight below, the surviving transport suggests a microscopic relation between the mean-field transition point~\cite{Iemini2018} and the bulk transport properties of the operator lattice.

For $j\rightarrow\infty$, the inter-rank hopping amplitudes scale as $t_\pm(k,q,j)\sim O(j)$, whereas the local damping terms scale as $\gamma(k,q)\sim O(1)$. After including the Kac scaling of the dissipator, the on-site decay vanishes, $\gamma(k,q) / j\rightarrow0,$ while the hopping amplitudes remain finite,
$ t_\pm(k,q,j) / j\sim O(1)$. Thus, in the thermodynamic limit, local damping vanishes but dissipative transport remains finite.

More explicitly, the BTC model retains the local hopping structure of
Eq.~(\ref{eq:non_hermitian_hopping_mapping}), with
$\gamma(k,q)/j\rightarrow0$ and
\begin{equation}
\begin{split}
t_-(k,q,j)
&=
\Bigl[(k-1)f(k,q)\Bigr]j+O(1),\\
t_+(k,q,j)
&=
-\Bigl[(k+2)f(k+1,q)\Bigr]j+O(1),
\end{split}
\end{equation}
where
\begin{equation}
f(k,q)=
\sqrt{
\frac{(k-q)(k+q)}
{(2k-1)(2k+1)}
}.
\end{equation}
The long lifetime of BTC oscillations is typically diagnosed through finite-size scaling of the Liouvillian gap~\cite{Iemini2018, Piccito2021, Nakanishi2023}. Our operator space representation reveals the microscopic origin of this behaviour: in the thermodynamic limit, the local damping contributions generated by the dissipator vanish as $j\rightarrow\infty$. Consequently, the thermodynamic BTC dynamics becomes a transport-dominated process in operator space, where operator weight propagates along the $k$ and $q$ directions. Note, this is fundamentally different from the model presented in Eq.~(\ref{eq:L_Jx_Jz}), in which case only the Hamiltonian-induced coherent transport remains, and each $k$-sector evolves independently.

The same operator space picture provides a microscopic connection to the BTC phase transition. In the thermodynamic limit the $(k,q)$ lattice becomes macroscopic, allowing the bulk transport properties to be analyzed for $1\ll k\ll j$. The coherent hopping generated by the Hamiltonian is
\begin{equation}
-i\Omega w_\pm(k,q)
=
-\frac{i\Omega}{2}\sqrt{k^2-q^2}
\sqrt{1+\frac{1}{k\pm q}},
\end{equation}
which away from the lattice boundaries, $q=\pm k$, reduces to
\begin{equation}
-i\Omega w_\pm(k,q)
\approx
-\frac{i\Omega}{2}\sqrt{k^2-q^2}.
\end{equation}
Similarly, the dissipative hopping amplitudes become
\begin{equation}
\frac{\Gamma}{2j}t_\pm(k,q)
\approx
\mp\frac{\Gamma}{4}\sqrt{k^2-q^2}.
\end{equation}
Hence, the bulk dynamics is governed by
\begin{equation}
\begin{split}
\frac{d}{dt}a_{k,q}
=&
\Bigl[
-i\Omega
(a_{k,q-1}+a_{k,q+1})
\\
&-\frac{\Gamma}{2}
(a_{k-1,q}-a_{k+1,q})
\Bigr]
g(k,q),
\label{eq:thermodynamic_limit_spherical_tensor_coeffs_evol}
\end{split}
\end{equation}
with
\begin{equation}
g(k,q)=\frac{1}{2}\sqrt{k^2-q^2}.
\end{equation}

To determine the corresponding bulk spectrum, we consider a locally uniform region around $(k_0,q_0)$ and
introduce the plane-wave ansatz~\cite{Sundaram1999, Silberstein2020, Berry1972}
\begin{equation}
a_{k,q}(t)
=
A(k_0,q_0)e^{\lambda t}
e^{ip_k (k-k_0)}e^{ip_q (q-q_0)},
\label{eq:plane_wave_ansatz}
\end{equation}
where $p_k$ and $p_q$ are conjugate momenta to the two operator space
directions. Substitution into Eq.~(\ref{eq:thermodynamic_limit_spherical_tensor_coeffs_evol}) yields
\begin{equation}
\lambda(p_k,p_q)
=
2i\Omega \, g(k_0, q_0)
\left(
r\sin(p_k) - \cos(p_q)
\right),
\end{equation}
where $r= \frac{\Gamma}{2\Omega}$.

Therefore, the mean-field critical point $\Gamma_c=2\Omega$ emerges directly from the bulk operator transport, marking the balance between coherent- and dissipative-dominated propagation.

\section{Discussion}
The operator space transport picture developed here suggests several extensions beyond the canonical collective spin models considered in this work. One immediate direction is to consider more general reservoir-engineered dissipators~\cite{Bai2021,Groszkowski2022}. In the spherical tensor representation, the irreducible tensor content of the jump operators constrains which tensor-rank sectors can be coupled. Higher-rank jump operators may therefore generate longer-ranged transport along the $k$-direction of the operator lattice, with the precise hopping structure determined by the irreducible tensor coupling rules and by cancellations between the different contributions to the Lindblad dissipator; see Appendix~\ref{app:dissipator_induced_k_hopping}. This raises the possibility of engineering the connectivity of the operator space lattice directly through the choice of dissipative channels.

A broader question is whether a similar transport-based description can be formulated for systems without permutation invariance. For generic dissipative spin chains, the spherical tensor basis could be replaced by a Pauli-string basis, which has been widely used to characterize operator spreading and growth in quantum circuits and many-body systems~\cite{Nahum2018,Keyserlingk2018,Khemani2018,Parker2019,Qi2023,Omanakuttan2023}. In such a representation, the relevant notion of transport would no longer be between collective tensor ranks, but between operators of different spatial support and complexity. Analogously, fermionic systems admit simple operator space descriptions in terms of Majorana operators, as exploited in Refs.~\cite{Prosen2007,Prosen2008}. It would therefore be interesting to investigate whether the symmetry-constrained transport picture introduced here admits a useful analogue in spatially resolved open many-body systems.

Such an extension would be particularly relevant for dissipative time crystals away from the idealized limit of perfect collectivity. Several works have investigated BTC-like behaviour in systems with spatial structure and long-range interactions~\cite{Passarelli2022,Wang2025,Russo2025}, but their descriptions have so far been limited to mean-field theory and cumulant expansions.

The role of non-unitality also raises several open questions. In the canonical BTC model, non-unital dissipation provides a source for traceless operator components and can dynamically populate oscillatory modes even when these are absent from the initial state. This offers a microscopic explanation for the initial state insensitivity observed in Refs.~\cite{Iemini2018,Mukherjee2024,nemeth2026topologicalboundarytimecrystal}, which is further discussed in Ref.~\cite{nemeth2026topologicalboundarytimecrystal}. It would be interesting to determine how general this mechanism is across other realizations of dissipative time crystals, and in particular whether the so-called quantum BTC regimes discussed in Refs.~\cite{Wang2025,Russo2025} admit a similar source-driven interpretation. More generally, this raises the question of whether non-unitality can be used as a diagnostic for identifying dissipative models capable of generating oscillatory dynamics independently of their initial traceless components.

In the thermodynamic limit, the BTC model can be further investigated as a bulk transport problem in operator space, in which the critical threshold $\Gamma_c=2\Omega$ emerges directly from the competition between coherent and dissipative propagation. This raises the broader question of whether operator space transport can provide a useful microscopic framework for describing dissipative phase transitions. In particular, it would be interesting to determine whether changes in the dynamical phase can be related to qualitative changes in the transport properties of operator space, such as transitions between localized and delocalized operator space eigenmodes. Establishing whether such a correspondence exists represents a natural direction for future work.

Finally, since our framework maps the collective spin dynamics directly onto a lattice model in operator space, it suggests connections with topological lattice systems, including free-fermion models. Given the progress in categorizing topology in such systems 
both from a Hermitian~\cite{Kruthoff2017,Bradlyn2017, Slager2013, Bouhon2020b,Po2017, Brouwer2023} as well as non-Hermitian Hamiltonian point of view~\cite{Hatano1996,Bender_2007,Kunst2021,Zhou2019,Kawabata2019,Borgnia2020,Okuma2020,Hatano1996}, this provides a direct area of research in topological features, for example, to address the robustness of BTC oscillations~\cite{nemeth2026topologicalboundarytimecrystal}.

\section{Conclusions}
In this work, we have developed a microscopic operator space framework for collective spin Lindbladians based on their weak symmetries and spherical tensor structure. By expressing the density operator in this tensor operator basis, we map the Lindbladian evolution onto a non-Hermitian hopping problem on an emergent operator space lattice. Within this representation, weak symmetries impose kinematic constraints on the allowed classes of operator space dynamics, which we interpret as selection rules on the permitted transport processes. The realized dynamical regime is then determined by the competition between the corresponding microscopic processes, in particular coherent intra-rank and dissipative inter-rank transport, including whether BTC oscillations emerge.

Beyond mean-field descriptions, our framework reveals how operator sectors beyond the collective spin triplet ($J_x, J_y$ and $J_z$) remain dynamically relevant and provides a microscopic explanation for the initial state insensitivity of BTC oscillations through non-unital operator space transport. Furthermore, the spherical tensor representation enables systematic analytical approaches, including perturbative descriptions of the extreme time crystal regime explored in Ref.~\cite{Nemethb2025}.

In the thermodynamic limit, we have shown that the BTC dynamics can be related to a bulk transport problem in operator space, where the mean-field phase transition threshold emerges from the competition between coherent and dissipative propagation. 

More broadly, our results suggest that dissipative many-body dynamics can be understood through the language of transport in operator space. This perspective provides a framework for studying non-equilibrium phases of open quantum systems and opens connections to broader concepts including operator spreading, non-Hermitian dynamics, and topology in quantum many-body systems.

\begin{acknowledgements}
D.N. acknowledges support from The University of Manchester under the Dean's Doctoral Scholarship. A.P. acknowledges support from the Leverhulme Trust under the grant agreement RPG-2023-253. R.-J.S. acknowledges funding from an EPSRC ERC underwrite Grant No.~EP/X025829/1.

\end{acknowledgements}

\appendix

\section{Adjoint Generators}
\label{sec:adjoint generators}
Under a symmetry transformation generated by $J_\alpha$, the density operator transforms as
\begin{equation}
\begin{split}
    & \rho \;\mapsto\; U \rho U^\dagger, \\ 
\end{split}
\end{equation}
where
\begin{equation}
    U = e^{-i\theta J_\alpha}
    \in \mathrm{SU}(2)
\end{equation}
in the spin-$j$ representation.
Expanding for infinitesimal $\theta$,
\begin{equation}
\begin{split}
    U \rho U^\dagger
    &= (\mathbf{1} - i\theta J_\alpha)\, \rho \, (\mathbf{1} + i\theta J_\alpha) \\
    &= \rho - i\theta [J_\alpha, \rho] + \mathcal{O}(\theta^2),
\end{split}
\end{equation}
which shows that the induced symmetry action on operators is generated by the adjoint (commutator) action,
\begin{equation}
   \mathcal{K}_\alpha[O] \equiv \mathrm{ad}_{J_\alpha}[O] = [J_\alpha, O].
\end{equation}

The operators $\{\mathcal{K}_\alpha\}$ therefore define a representation of the $\mathfrak{su}(2)$ algebra on the operator space $\mathcal{B}(\mathcal{H}_j)$. In direct analogy with the closed-system case, one may construct the corresponding Casimir superoperator,
\begin{equation}
    \mathcal{K}^2 = \mathcal{K}_x^2 + \mathcal{K}_y^2 + \mathcal{K}_z^2.
\end{equation}
Simultaneous diagonalization of $\mathcal{K}^2$ and $\mathcal{K}_z$ yields the operator space quantum numbers $(k,q)$, which label irreducible multiplets. For each fixed $k$, the allowed values of $q$ are
\begin{equation}
    q = -k, -k+1, \ldots, k,
\end{equation}
forming a $(2k+1)$-dimensional irrep.

\section{Spherical Tensor Operators}
\label{app:spherical_tensors}
In this appendix, we summarize standard results on spherical tensor operators used throughout the main text.

A spherical tensor operator of rank $k$ and component $q$, denoted $T^k_q$, is defined through its transformation under the adjoint action of the collective spin generators~\cite{Varshalovich1988, Hecht2000},
\begin{equation}
\begin{split}
\mathcal{K}_z[T^k_q] &\equiv [J_z, T^k_q] = q\, T^k_q, \\
\mathcal{K}_\pm[T^k_q] &\equiv [J_\pm, T^k_q]
= \sqrt{k(k+1)-q(q\pm1)}\, T^k_{q\pm1}.
\end{split}
\label{eq:spherical_tensor_defns}
\end{equation}

For a collective spin system with fixed total spin $j$, the operators $T^k_q$ furnish an explicit basis for the irreducible representations appearing in the decomposition
\begin{equation}
j \otimes j^\ast = \bigoplus_{k=0}^{2j} k.
\end{equation}
An explicit construction in terms of angular momentum eigenstates $\{\ket{j\, m}\}_{m=-j}^{j}$ is given by
\begin{equation}
T^k_q
=
\sum_{m,m'=-j}^{j} (-1)^{j-m}
\langle j\, m'; j\,{-}m \vert k\, q \rangle
\, \ket{j\, m'} \bra{j\, m},
\end{equation}
where $\langle j\, m'; j\,{-}m \vert k\, q \rangle$ are Clebsch--Gordan coefficients~\cite{Blum2012}.

The spherical tensor operators are chosen to be orthonormal with respect to the Hilbert--Schmidt inner product~\cite{Blum2012},
\begin{equation}
\mathrm{Tr}\!\left[(T^k_q)^\dagger T^{k'}_{q'}\right]
=
\delta_{k k'}\,\delta_{q q'},
\label{eq:spherical_tensor_orthonormality}
\end{equation}
and therefore form a complete orthonormal basis for operators acting within the spin-$j$ sector.

Using the orthonormality of spherical tensor operators, one can show that all tensors with $(k \geq 1)$ are traceless. Recall that the identity operator corresponds to the unique rank-zero tensor $T^0_0 = \frac{1}{\sqrt{2j+1}}\,\mathbf{1} $. Then,
\begin{equation}
\begin{split}
    \Tr(T^k_q)
    &= \Tr\!\left[(\mathbf 1)^\dagger T^k_q\right] \\
    &= {\sqrt{2j+1}} \,\Tr\!\left[(T^0_0)^\dagger T^k_q\right] \\
    &= {\sqrt{2j+1}} \,\delta_{k0}\,\delta_{q0} \\
    &= 0 \qquad \text{for } k \geq 1.
\end{split}
\label{eq:traceless_tensors}
\end{equation}

\vspace{0.3em}

Next, we will outline the coupling rules obeyed by irreducible tensors with arbitrary normalizations and later specialize to the case where both tensors describe the state of a spin-$j$ object.

Irreducible tensor operators obey angular momentum coupling rules analogous to those for state vectors. In general, the direct product of two irreducible tensor operators decomposes as
\begin{equation}
V^{k_1}_{q_1} \, U^{k_2}_{q_2}
=
\sum_{k = |k_1 - k_2|}^{k_1 + k_2}
\langle k_1\, q_1; k_2\, q_2 \vert k\, q \rangle
\, W^k_q,
\label{eq:direct_product_of_irreducible_tensors}
\end{equation}
where the allowed values of $k$ satisfy the triangle rule
\[
k = |k_1 - k_2|, |k_1 - k_2| + 1, \ldots, k_1 + k_2,
\qquad q = q_1 + q_2.
\]
Thus, the composition of irreducible tensor operators follows the same algebra as the addition of angular momenta. This construction extends straightforwardly to products of three or more tensors by successive application of Eq.~(\ref{eq:direct_product_of_irreducible_tensors}).

For the irreducible tensors considered here, $T^{k_1}_{q_1}(j)$ and $T^{k_2}_{q_2}(j)$, both irreducible tensors describe the polarization of an underlying spin-$j$ particle~\cite{Varshalovich1988} and satisfy the following relations
\begin{widetext}
    \begin{equation}
    T^{k_1}_{q_1}(j) T^{k_2}_{q_2}(j) = \sum_{K} (-1)^{2j+K} \sqrt{(2k_1+1)(2k_2+1)} 
    \begin{Bmatrix}
         k_1 & k_2 & K \\
        j & j & j \\
    \end{Bmatrix}
    \langle k_1 q_1; k_2 q_2\vert K Q\rangle \, T^K_Q(j),
    \label{eq:polarization_ops_tensor_rule}
\end{equation}
and
\begin{equation}
       \{T^{k_1}_{q_1}(j), T^{k_2}_{q_2}(j)\} = \sum_{K} (-1)^{2j+K} [1 + (-1)^{k_1+k_2+K} ]\sqrt{(2k_1+1)(2k_2+1)} 
    \begin{Bmatrix}
         k_1 & k_2 & K \\
        j & j & j \\
    \end{Bmatrix}
    \langle k_1 q_1; k_2 q_2\vert K Q\rangle \, T^K_Q(j),   
    \label{eq:polarization_ops_anticommutator_rule}
    \end{equation}
\end{widetext}
where
\begin{equation}
    \begin{Bmatrix}
         k_1 & k_2 & K \\
        j & j & j \\
    \end{Bmatrix}
\end{equation}
is a $6j$ symbol that enforces the triangle rule $K= \vert k_1-k_2\vert, \ldots, k_1+k_2$. Here, $\langle k_1 q_1; k_2 q_2\vert K Q\rangle$ is a Clebsch--Gordan coefficient. Importantly, the $6j$ symbols possess a high degree of permutation symmetry: they are invariant under the interchange of any two columns, as well as under the simultaneous interchange of upper and lower entries in any two columns. Clebsch–Gordan coefficients, however, are sensitive to the ordering of the coupled tensors. Permuting their arguments generally introduces phase factors and therefore the chosen coupling convention must be applied consistently throughout the derivation to preserve the correct sign structure. See Ref.~\cite{Varshalovich1988} for further details.

\section{General Form of Evolution Equation in the Spherical Tensor Basis}
\label{app:general_evol_eqn_in_tensor_basis}
Eq.~(\ref{eq:liouvillian_spherical_tensor_basis_general}) follows directly by expanding the density operator in the spherical tensor basis [Eq.~(\ref{eq:spherical_tensor_basis})] and inserting this into the master equation $\tfrac{d}{dt}\rho = \mathcal{L}\rho$. Explicitly,
\begin{equation}
\begin{split}
    \frac{d}{dt} \Bigl (\sum_{k'q'} a_{k', q'} (t) \,  T^{k'}_{q'}\Bigr)
    &= \mathcal L \Bigl( \sum_{k'q'} a_{k', q'} (t) \,  T^{k'}_{q'} \Bigr) \\
    \implies \sum_{k'q'} \Bigl(\frac{d}{dt} a_{k', q'} (t) \Bigr)\, T^{k'}_{q'}
    &= \sum_{k'q'} a_{k', q'} (t) \, \mathcal L \,  T^{k'}_{q'} .
\end{split}
\end{equation}

Projecting onto a given spherical tensor mode by multiplying from the left with $(T^k_q)^\dagger$ and taking the trace yields
\begin{equation}
\begin{split}
  & \sum_{k'q'} \bigl(\tfrac{d}{dt}a_{k', q'} (t)\bigr) \Tr \Bigl ((T^k_q)^\dagger T^{k'}_{q'} \Bigr) \\
  & = \sum_{k'q'} a_{k', q'} (t) \, \Tr \Bigl((T^k_q)^\dagger\, \mathcal L \,  T^{k'}_{q'} \Bigr).\\
\end{split}
\end{equation}

Using the orthonormality relation [Eq.~(\ref{eq:spherical_tensor_orthonormality})], we obtain
\begin{equation}
    \frac{d}{dt} a_{k,q}(t) = \sum_{k'q'} \mathcal L_{(k,q)(k', q')} \, a_{k',q'}(t),
\end{equation}
where the matrix elements of the Liouvillian in the spherical tensor basis are given by
\begin{equation}
    \mathcal L_{(k,q)(k', q')} \equiv \Tr \Bigl[(T^k_q)^\dagger\, \mathcal L \,  T^{k'}_{q'} \Bigr].
\end{equation}

\section{Mapping to a Non-Hermitian Hopping Model (Collective Precession)}
\label{app:collective_precession_mapping}
In this section we derive Eq.~(\ref{eq:collective_precession_hopping_model}) by expressing the Liouvillian $\mathcal L$ in the spherical tensor basis $\{T^k_q\}$ [cf.~Eq.~(\ref{eq:liouvillian_kq_matrix_elements})].

The unitary contribution may be written in terms of the adjoint generators as $\mathcal K_x[\cdot] = [J_x,\cdot]$, using Eq.~(\ref{eq:adjoint_generator_definition}). For Hermitian jump operators, the dissipator admits the double-commutator form
\begin{equation}
    J_z T^k_q J_z - \tfrac{1}{2}\{J_z^2, T^k_q \}
    = - \tfrac{1}{2}[J_z, [J_z, T^k_q ]].
\end{equation}
Using the defining relations of spherical tensors [Eq.~(\ref{eq:spherical_tensor_defns})], we obtain
\begin{equation}
    [J_z, [J_z, T^k_q]]
    = \mathcal K_z\!\left[\mathcal K_z[T^k_q]\right]
    = q^2 \, T^k_q,
\end{equation}
which shows that the dissipator contributes a diagonal term proportional to $q^2$.

The Liouvillian therefore takes the compact form
\begin{equation}
    \mathcal L = -i\Omega \, \mathcal K_x - \frac{\Gamma}{4j}\,\mathcal K_z^2.
\end{equation}

We now evaluate the matrix elements in Eq.~(\ref{eq:liouvillian_kq_matrix_elements}). The unitary contribution reads
\begin{equation}
\begin{split}
\mathrm{Tr}\!\Bigl[(T^k_q)^\dagger \mathcal K_x T^{k'}_{q'} \Bigr]
= {} & \tfrac{1}{2} \mathrm{Tr}\!\Bigl[(T^k_q)^\dagger [J_+, T^{k'}_{q'}]\Bigr] \\
& + \tfrac{1}{2} \mathrm{Tr}\!\Bigl[(T^k_q)^\dagger [J_-, T^{k'}_{q'}]\Bigr],
\end{split}
\end{equation}
where we used $J_x = \tfrac{1}{2}(J_+ + J_-)$. Applying Eq.~(\ref{eq:spherical_tensor_defns}) yields
\begin{equation}
\begin{split}
& \tfrac{1}{2}\sqrt{k'(k'+1) - q'(q'+1)}\;
\mathrm{Tr}\!\Bigl[(T^k_q)^\dagger T^{k'}_{q'+1} \Bigr] \\
+\, & \tfrac{1}{2}\sqrt{k'(k'+1) - q'(q'-1)}\;
\mathrm{Tr}\!\Bigl[(T^k_q)^\dagger T^{k'}_{q'-1} \Bigr].
\end{split}
\end{equation}
Using the orthonormality relation [Eq.~(\ref{eq:spherical_tensor_orthonormality})], this reduces to
\begin{equation}
    w_+(k',q')\,\delta_{k,k'}\delta_{q,q'+1}
    + w_-(k',q')\,\delta_{k,k'}\delta_{q,q'-1},
\end{equation}
where we defined
\[
w_{\pm}(k,q)=\tfrac{1}{2}\sqrt{k(k+1)-q(q\pm1)}.
\]

Including the dissipative contribution, we obtain the Liouvillian matrix elements
\begin{equation}
\begin{split}
\mathcal{L}_{(k,q)(k',q')} 
= {} & -i \Omega \Bigl[
    w_+(k',q')\,\delta_{k,k'}\delta_{q,q'+1} \\
& \qquad\;\; + w_-(k',q')\,\delta_{k,k'}\delta_{q,q'-1}
\Bigr] \\
& - \frac{\Gamma}{4j}\, q^2 \,\delta_{k,k'} \delta_{q,q'}.
\end{split}
\end{equation}

Substituting this result into Eq.~(\ref{eq:liouvillian_spherical_tensor_basis_general}) yields the hopping model in Eq.~(\ref{eq:collective_precession_hopping_model}).

\section{Eigensystem in the $k=1$ Subspace (Collective Precession)}
\label{app:q_mixing_eigensystem}
We now derive the equations of motion for the expansion coefficients $a_{k,q}(t)$ [cf.~Eq.~(\ref{eq:odes_for_exp_coeffs_q_mixing})]. Substituting into the master equation yields
\begin{equation}
\begin{split}
    \frac{d}{dt}a_{k,q}(t)
    =
    &- \frac{i \Omega}{2}
    \Bigl[
    \sqrt{k(k+1) - (q-1)q} \; a_{k,\,q-1}(t) \\
    &\qquad\quad
    + \sqrt{k(k+1) - q(q+1)} \; a_{k,\,q+1}(t)
    \Bigr] \\
    &- \frac{\Gamma}{4j} \, q^2 \, a_{k,q}(t).
\end{split}
\end{equation}

\vspace{0.3em}

To diagonalize this system, we focus on the $k=1$ sector and introduce symmetric and antisymmetric combinations,
\begin{equation}
    a_{\pm}(t) = \tfrac{1}{\sqrt{2}}\bigl(a_{1,1}(t) \pm a_{1,-1}(t)\bigr).
\end{equation}
In this basis, the antisymmetric mode decouples,
\begin{equation}
    \frac{d}{dt} a_-(t) = - \frac{\Gamma}{4j} \, a_-(t),
\end{equation}
with solution
\begin{equation}
    a_-(t) = a_-(0)\, e^{-\tfrac{\Gamma}{4j}t}.
\end{equation}

The remaining components $(a_{1,0}, a_+)$ satisfy
\begin{equation}
    \frac{d}{dt}
    \begin{pmatrix}
        a_{1,0}(t) \\
        a_+(t)
    \end{pmatrix}
    =
    \begin{pmatrix}
        0 & -i\Omega \\
        -i\Omega & -\tfrac{\Gamma}{4j}
    \end{pmatrix}
    \begin{pmatrix}
        a_{1,0}(t) \\
        a_+(t)
    \end{pmatrix}.
\end{equation}
The corresponding eigenvalues are
\begin{equation}
    \lambda_{\pm}
    =
    -\frac{\Gamma}{8j}
    \pm i \Omega \sqrt{1 - \left(\frac{\kappa}{2}\right)^2},
\end{equation}
where $\kappa \equiv \Gamma/(4j\Omega)$.

\vspace{0.3em}

To connect to physical observables, we use the relations between spherical tensors and spin operators (following Ref.~\cite{Blum2012}),
\begin{equation}
    T^1_0 = N_1(j)\, J_z,
    \qquad
    T^1_{\pm1} = \mp\frac{N_1(j)}{\sqrt{2}}\, J_{\pm},
\end{equation}
with normalization
\begin{equation}
    N_1(j) = \left[\frac{3}{(2j+1)(j+1)j}\right]^{1/2}.
\end{equation}
This implies that the symmetric and antisymmetric combinations correspond to
\begin{equation}
\begin{split}
    T_{+} &= - \sqrt{2}\, N_1(j)\, i J_y, \\
    T_{-} &= - \sqrt{2}\, N_1(j)\, J_x.
\end{split}
\end{equation}

\vspace{0.3em}

We therefore conclude that $J_x$ decouples and exhibits purely exponential decay with rate $\Gamma/(4j)$, while $J_y$ and $J_z$ form a coupled pair undergoing damped oscillations with decay rate $\Gamma/(8j)$ and frequency $\Omega \sqrt{1 - (\kappa/2)^2}$. This matches the results presented in Ref.~\cite{Nemethb2025}.

\section{Dissipator-Induced $k$-hopping (BTC)}
\label{app:dissipator_induced_k_hopping}
Here, we show that the dissipator in Eq.~(\ref{eq:canonical_BTC}) couples spherical tensor operators with different tensor rank $k$. A simple way to see this is to evaluate its action on low-rank tensors.

The $k=0$ sector is one-dimensional, with $T^0_0 = \mathbf{1}/\sqrt{2j+1}$. Acting with the dissipator yields
\begin{equation}
\begin{split}
    \mathcal D[\mathbf{1}]
    &=  \frac{\Gamma}{2j}\Bigl(J_- \mathbf{1} J_+ - \tfrac{1}{2}(J_+ J_- \mathbf{1} + \mathbf{1} J_+ J_-) \Bigr)\\
    &= \frac{\Gamma}{2j}\Bigl( J_- J_+ - J_+ J_- \Bigr)\\
    &= \frac{\Gamma}{2j}[J_-, J_+] \\
    &= -\frac{\Gamma}{j}J_z \;\propto\; T^1_0,
\end{split}
\end{equation}
showing that the identity operator (rank $k=0$) is mapped to a rank-$1$ tensor.

Next, consider the action on $J_z \propto T^1_0$. A straightforward calculation gives
\begin{equation}
\begin{split}
    \mathcal D[J_z]
    &= \frac{\Gamma}{2j} \Bigl(J_- J_z J_+ - \tfrac{1}{2}(J_+ J_- J_z + J_z J_+ J_-) \Bigr)\\
    &= \frac{\Gamma}{2j}\Bigl(J^2 - 3 J_z^2- J_z\Bigr) ,
\end{split}
\end{equation}
where we have used $[J_z, J_\pm] = \pm J_\pm$ and $[J^2, J_z] = 0$. This expression can be decomposed into spherical tensor components as
\begin{equation}
    \mathcal D[J_z]
    =\frac{\Gamma}{2j}\Bigl(
    -J_z - (3J_z^2 - J^2)\Bigr)
\end{equation}
Here, $J_z \propto T^1_0$ and $(3J_z^2 - J^2) \propto T^2_0$. Thus, the dissipator maps a rank-$1$ tensor onto a superposition of tensors with ranks $k=1$ and $k=2$.

These examples demonstrate that the dissipator preserves the quantum number $q$, while it can induce nearest-neighbor coupling in $k$. In particular, it generates transitions $k \to k\pm1$, implying that
\begin{equation}
    [\mathcal D, \mathcal K^2] \neq 0.
\end{equation}

This structure admits a representation-theoretic interpretation. Since the jump operators $J_\pm$ transform as rank-$1$ irreducible tensors, the product $J_- T^k_q$ corresponds to the coupling of tensors with ranks $1$ and $k$. By angular momentum addition, this decomposes into irreducible tensors with ranks $K = k-1, k, k+1$. Acting subsequently with $J_+$ produces tensors with ranks $K' = k-2, k-1, k, k+1, k+2$, while the component remains fixed at $q$, since $Q' = -1 + q + 1 = q$. 

However, for the specific Lindblad structure in Eq.~(\ref{eq:canonical_BTC}), cancellations between the jump term and the anti-commutator remove the $K' = k \pm 2$ contributions. As a result, the dissipator couples only the neighboring ranks $k-1$, $k$, and $k+1$.  We prove this explicitly below.

We first note that
\begin{equation}
\begin{split}
        J_- T^k_q J_+ -\frac{1}{2}\{J_+J_-, T^k_q\} = & \frac{1}{2}[J_-, T^k_q J_+] +\frac{1}{2}[J_-T^k_q, J_+],
\end{split}
\end{equation}
which can be further simplified to
\begin{equation}
    \begin{split}
        \frac{1}{2}[J_-, T^k_q]J_+ +\frac{1}{2}J_-[T^k_q, J_+] - \{J_z, T^k_q\}
    \end{split}
    \label{eq:dissipator_mid_big_calculation}
\end{equation}
using $[J_-, J_+]=- 2J_z$. The commutator terms can be simplified via Eq.~(\ref{eq:spherical_tensor_defns}) to obtain terms proportional to $T^k_{q\pm1}$. Then, $T^k_{q-1} J_+$ and $J_-T^{k}_{q+1}$ may be evaluated using Eq.~(\ref{eq:polarization_ops_tensor_rule}) and the anticommutator via Eq.~(\ref{eq:polarization_ops_anticommutator_rule}). For each of these three terms, the triangle rule enforces $K= k-1, k, k+1$, which already demonstrates that only nearest-neighbor and on-site terms remain. We can evaluate the coefficients explicitly by noting that
\begin{widetext}
\begin{equation}
    \begin{Bmatrix}
        1 & k & k \\
        j & j & j \\
    \end{Bmatrix}
    = (-1)^{-2j-k+1} \sqrt{\frac{k(k+1)}{4j(j+1)(2j+1)(2k+1)}}
\end{equation}
for $1\leq k \leq 2j$ and $0$ for $k=0$,
\begin{equation}
    \begin{Bmatrix}
        1 & k & k-1 \\
        j & j & j \\
    \end{Bmatrix}
    = (-1)^{-2j-k} \sqrt{\frac{k\Bigl[(2j+1)^2- k^2\Bigr]}{4j(j+1)(2j+1)(2k-1)(2k+1)}},
\end{equation}
for $1\leq k \leq 2j$ and $0$ for $k=0$ and
\begin{equation}
    \begin{Bmatrix}
        1 & k & k+1 \\
        j & j & j \\
    \end{Bmatrix}
    = (-1)^{-2j-k+1} \sqrt{\frac{(k+1)\Bigl[(2j+1)^2- (k+1)^2\Bigr]}{4j(j+1)(2j+1)(2k+1)(2k+3)}},
\end{equation}
for $0\leq k \leq 2j-1$ and $0$ for $k=2j$.
\end{widetext}
The $6j$ symbols are implemented in various numerical software, for example in Mathematica via the \texttt{SixJSymbol} function, or can be evaluated analytically from Racah coefficients given in Ref.~\cite{Varshalovich1988}. These expressions appear in all three terms of Eq.~(\ref{eq:dissipator_mid_big_calculation}), which yield $K=k-1, k, k+1$ via the triangle rule.

Combining these results we obtain
\begin{equation}
\begin{split}
     \mathcal{D}[T^k_q] = &\frac{\Gamma}{2j}\Bigl( t_-(k,q,j)T^{k-1}_q + t_+(k,q,j)T^{k+1}_q \\
     & -\gamma(k,q,j) T^k_q\Bigr).
\end{split}
\end{equation}
Here, the on-site damping factors are independent of $j$ such that
\begin{equation}
    \gamma(k,q,j)= \gamma(k,q) = \frac{1}{2}\Bigl[k(k+1)-q^2\Bigr],
\end{equation}
for $0 \leq k \leq 2j$ and $-k\leq q \leq +k$. The hopping amplitude $t_-(k,q,j)$ is given by
\begin{equation}
    t_-(k,q,j) = \frac{(k-1)}{2}\sqrt{\frac{\Bigl[(2j+1)^2 -k^2\Bigr]\Bigl(k^2-q^2\Bigr)}{(2k-1)(2k+1)}},
\end{equation}
for $1\leq k \leq 2j$ and $-k\leq q \leq +k$ and it vanishes for $k=1$, reflecting trace preservation. Furthermore, $t_-(k,q, j)\propto\sqrt{k^2-q^2}$, implying that hopping out of the extremal multiplet states $q=\pm k$ is forbidden. Geometrically, this is because the neighbouring rank-($k-1$) multiplet only supports $\vert q\vert \le k-1$, so no destination states $(k-1,\pm k)$ exist.

Similarly,
\begin{equation}
\begin{split}
    & t_+(k,q,j) = -\frac{(k+2)}{2}\,  S(k,q, j),\\
    & S(k,q, j)= \sqrt{\frac{(2j-k)(2j+k+2)\Bigl[(k+1)^2 - q^2\Bigr]}{(2k+1)(2k+3)}},\\
\end{split}
\end{equation}
is valid for $0 \leq k \leq 2j -1$ and $-k \leq q \leq +k$. This hopping amplitude vanishes for $k=2j$ since it corresponds to the upper edge of the lattice along the $k$-coordinate.

In practice, one can also extract these coefficients numerically from the matrix elements of the Liouvillian in the $\{T^k_q\}$ basis. This basis can either be constructed from angular momentum states or by finding the simultaneous eigenvectors of the vectorized superoperators $\mathbb K^2$ and $\mathbb K_z$.

In the thermodynamic limit $j\to\infty$, one can expand $t_\pm(k,q,j)$ in powers of $j$ with the leading order contributions being $O(j)$ followed by $O(1)$, $O(j^{-1})$ etc.

We note that nearest-neighbor hoppings arising from dissipators with collective jump operators $J_\pm$ have also been identified in Ref.~\cite{Ribeiro2019}. However, that work is restricted to models with a conserved charge, in contrast to the present setting.

\section{Projectors for $\mathcal K^2$ Eigenspaces}
\label{sec:projectors}
In this section, we provide further details on the projectors used in the main text.

In the vectorized Liouville space, the spherical tensor basis $\{\ket{k\, q}\rangle\}$ satisfies the completeness relation
\begin{equation}
\begin{split}
    \mathbb I
    &= \sum_{k=0}^{2j} \sum_{q=-k}^{+k} \ket{k\, q}\rangle \langle\bra{k\, q} \\
    &= \sum_{k=0}^{2j}
    \underbrace{\Bigl(\sum_{q=-k}^{+k} \ket{k\, q}\rangle \langle\bra{k\, q} \Bigr)}_{\mathbb P^{(k)}} \\
    &= \sum_{k=0}^{2j} \mathbb P^{(k)}.
\end{split}
\end{equation}
where $\mathbb P^{(k)} = \bigl(\mathbb P^{(k)}\bigr)^\dagger$ projects onto the eigenspace of $\mathbb K^2$ with eigenvalue $k(k+1)$.

\vspace{0.3em}

Using these projectors, the Liouvillian can be decomposed into blocks connecting different tensor sectors,
\begin{equation}
\begin{split}
    \mathbb D^{(k' \to k)}
    &\equiv \mathbb P^{(k)} \, \mathbb D \, \mathbb P^{(k')} \\
    &= \sum_{q=-k}^{+k} \sum_{q'=-k'}^{+k'}
    \ket{k\, q}\rangle
    \langle\bra{k\, q}\, \mathbb D \, \ket{k'\, q'}\rangle
    \langle \bra{k'\, q'} \\
    &= \sum_{q=-k}^{+k} \sum_{q'=-k'}^{+k'}
    \ket{k\, q}\rangle
    \bigl(\mathbb D^{(k' \to k)}\bigr)_{q q'}
    \langle \bra{k'\, q'}.
\end{split}
\end{equation}
This operator describes the transfer of operator weight from the $k'$ sector to the $k$ sector under the action of $\mathbb D$.

\vspace{0.3em}

To quantify the strength of this coupling, we define
\begin{equation}
\begin{split}
    C^{(k' \to k)}
    &\equiv \sqrt{\mathrm{Tr}\!\left[
    \bigl(\mathbb D^{(k' \to k)}\bigr)^\dagger
    \mathbb D^{(k' \to k)}
    \right]} \\
    &= \sqrt{\sum_{q,q'} \left| \bigl(\mathbb D^{(k' \to k)}\bigr)_{q q'} \right|^2},
\end{split}
\end{equation}
which corresponds to the Frobenius norm of the block $\mathbb D^{(k' \to k)}$.

\end{document}